# A population genetic interpretation of GWAS findings for human quantitative traits

Yuval B. Simons[a,1], Kevin Bullaughey[b], Richard R. Hudson[b], and Guy Sella[a,1]

[a] Department of Biological Sciences, Columbia University, New York, NY 10027;
[b] Department of Ecology & Evolution, University of Chicago, Chicago, Illinois 60637;
[1] To whom correspondence should be addressed: yuval.simons@columbia.edu, gs2747@columbia.edu

## Abstract

Genome-wide association studies (GWAS) in humans are revealing the genetic architecture of biomedical and anthropomorphic traits, i.e., the frequencies and effect sizes of variants that contribute to heritable variation in a trait. To interpret these findings, we need to understand how genetic architecture is shaped by basic population genetics processes—notably, by mutation, natural selection and genetic drift. Because many quantitative traits are subject to stabilizing selection and genetic variation that affects one trait often affects many others, we model the genetic architecture of a focal trait that arises under stabilizing selection in a multi-dimensional trait space. We solve the model for the phenotypic distribution and allelic dynamics at steady state and derive robust, closed form solutions for summary statistics of the genetic architecture. Our results provide a simple interpretation for missing heritability and why it varies among traits. They also predict that the distribution of variances contributed by loci identified in GWAS is well approximated by a simple functional form that depends on a single parameter: the expected contribution to genetic variance of a strongly selected site affecting the trait. We test this prediction against the results of GWAS for height and body mass index (BMI) and find that it fits the data well, allowing us to make inferences about the degree of pleiotropy and mutational target size for these traits. Our findings help to explain why the GWAS for height explains more of the heritable variance than similarly-sized GWAS for BMI, and to predict the increase in explained heritability with study sample size. Considering the demographic history of European populations, in which these GWAS were performed, we further find that most of the associations they identified likely involve mutations that arose during the out of Africa bottleneck at sites with selection coefficients around $s = 10^{-3}$.

Much of the phenotypic variation in human populations, including variation in morphological, life history and biomedical traits, is "quantitative", in the sense that heritable variation in the trait is largely due to small contributions from many genetic variants segregating in the population (1, 2). Quantitative traits have been studied since the birth of biometrics over a century ago (1-3), but only in the past decades have technological advances made it possible to systematically dissect their genetic basis (4-6). Notably, since 2007, genome-wide association studies (GWAS) in humans have led to the identification of many thousands of variants reproducibly associated with hundreds of quantitative traits, including susceptibility to a wide variety of diseases (4). While still ongoing, these studies already provide important insights into the genetic architecture of quantitative traits, i.e., the number of variants that contribute to heritable variation, as well as their frequencies and effect sizes.

Perhaps the most striking observation to emerge from these studies is that, despite the large sample size of many GWAS, all variants significantly associated with any given trait typically account for less (often much less) than 25% of the narrow sense heritability (4, 7, 8, but see 9). (Henceforth, we use "heritability" to refer to narrow sense heritability.) While many factors have been hypothesized to contribute to the "missing heritability" (7, 8, 10-14), the most straightforward explanation and the emerging consensus is that much of the heritable variation derives from variants with frequencies that are too low or effect sizes that are too small for current studies to detect. Comparisons among traits also suggest that there are substantial differences in architectures. For example, recent meta-analyses GWAS uncovered seven times as many variants for height (697) than for body mass index (97), and together the variants for height account for more than four times the heritable variance for body mass index (~20% vs. ~3-5%, respectively), despite comparable sample sizes (15, 16).

These first glimpses underscore the need for theory that relates the findings emerging from GWAS with the evolutionary processes that shape genetic architectures. Such theory would help to interpret the "missing heritability" (17-20) and to explain why architecture differs among traits. It may also allow us to use GWAS findings to make inferences about underlying evolutionary parameters, helping to answer enduring questions about the processes that maintain phenotypic variation in quantitative traits (5, 21).

Development of such theory can be guided by empirical observations and first principles considerations. New mutations affecting a trait arise at a rate that depends on its "mutational target size" (i.e., the number of sites at which a mutation would affect the trait). Once they arise, the trajectories of variants through the population are determined by the interplay between genetic drift, demographic processes, and natural selection acting on them. These processes determine the number and frequencies of segregating variants underlying variation in the trait. The genetic architecture further depends on the relationship between the selection on variants and their effects on the trait. Notably, selection on variants depends not only on their effect on the focal trait but also on their pleiotropic effects on other traits. We therefore expect both direct and pleiotropic selection to shape the joint distribution of allele frequencies and effect sizes.

Multiple lines of evidence suggest that many quantitative traits are subject to stabilizing selection, i.e., selection favoring an intermediate trait value (5, 22-26). For instance, a decline in fitness components (e.g., viability and fecundity) is observed with displacement from mean values for a variety of traits in human populations (27-29), in other species in the wild (30, 31) and in experimental manipulations (30, 32). While less is known about complex diseases, they may often reflect large deviations of an underlying continuous trait from an optimal value, with these continuous traits subject to directional (purifying) selection in some cases and to stabilizing selection in others. What remains unclear is the extent to which stabilizing selection is acting directly on variation in a given trait or is "apparent", i.e., results from pleiotropic effects of this variation on other traits.

Other lines of evidence suggest that pleiotropy is pervasive. For one, theoretical considerations about the variance in fitness in natural populations and its accompanying genetic load suggest that only a moderate number of independent traits can be effectively selected on at once (33). Thus, the aforementioned relationships between the value of a focal trait and fitness are likely heavily affected by the pleiotropic effects of genetic variation on other traits (25, 33-35). Second, many of the variants detected in human GWAS have been found to be associated with more than one trait (36-40). For example, a recent analysis of GWAS revealed that variants that delay the age of menarche in women tend to delay the age of voice drop in men, decrease body mass index, increase adult height, and decrease risk of male pattern baldness (36). More generally, the extent of pleiotropy revealed by GWAS appears to be increasing rapidly with improvements in power and methodology (36, 41-44). These considerations and others (44, 45) point to the general importance of pleiotropic selection on quantitative genetic variation.

The discoveries emerging from human GWAS further suggest that genetic variance is dominated by additive contributions from numerous variants with small effect sizes. Dominance and epistasis may be common among newly arising mutations of large effect (e.g., 46, 47-50), but both theory and data suggest that they play a minor role in shaping quantitative genetic variation within populations (e.g., 9, 51, 52-55). Indeed, for many traits, most or all of the heritability explained in GWAS arises from the additive contribution of variants with squared effect sizes that are substantially smaller than the total genetic variance (e.g., 15, 16, 56, 57). Moreover, statistical quantifications of the total genetic variance tagged by genotyping (i.e., not only due to the genome-wide significant associations) suggest that such contributions may account for most of the heritable variance in many traits (e.g., 9, 58-60). Finally, considerable efforts to detect epistatic interactions in human GWAS have, by and large, come up empty-handed (9, 55, 61), with few counter-examples mostly involving variants in the MHC region (52, 55, 62, 63, but see 64). Thus, while the discovery of epistatic interactions may be somewhat limited by statistical power (55), theory and current evidence suggest that that non-additive interactions play a minor role in shaping human quantitative genetic variation. Motivated by these considerations, we model how direct and pleiotropic stabilizing selection shape the genetic architecture of continuous, quantitative traits by considering additive variants with small effects and assuming that together they account for most of the heritable variance.

To date, there has been relatively little theoretical work relating population genetics processes with the results emerging from GWAS. Moreover, the few existing models have reached divergent predictions about genetic architecture, largely because they make different assumptions about the effects of pleiotropy. Focusing on disease susceptibility, Pritchard (19) considered the "purely pleiotropic" extreme, in which selection on variants is independent of their effect on the trait being considered. In this case, we expect the largest contribution to genetic variance in a trait to come from mutations that have large effect sizes but are also weakly selected or neutral, allowing them to ascend to relatively high frequencies. Other studies considered the opposite extreme, in which selection on variants stems entirely from their effect on the trait under consideration (26, 65-69), and have shown that the greatest contribution to genetic variance would arise from strongly selected mutations (66, 67) (we return to this case below).

In practice, we expect most traits to fall somewhere in between these two extremes. While there are compelling reasons to believe that quantitative genetic variation is highly pleiotropic, the effects of variants on different traits are likely to be correlated. Thus, even if a given trait is not subject to selection, variants that have a large effect on it will also tend to have larger effects on traits that are under selection (e.g., by causing large perturbation to pathways that affect multiple traits; 35, 44). Motivated by such considerations, Eyre-Walker (2010), Keightley and Hill (1990), and Caballero et al. (2015) considered models in which the correlation between the strength of selection on an allele and its effect size can vary between the purely pleiotropic and direct selection extremes. These models diverge in their predictions about architecture, however. Assuming, as seems plausible, an intermediate correlation between the strength of selection and effect size, Eyre-Walker finds that genetic variance should be dominated by strongly selected mutations (20), whereas Keightley & Hill and Caballero et al. conclude that the greatest contribution should arise from weakly selected ones (18, 70). Their conclusions differ because of how they chose to model the relationship between selection and effect size, a choice based largely on mathematical convenience. We approach this problem by explicitly modeling stabilizing selection on multiple traits, thereby learning, rather than assuming, the relationship between selection and effect sizes.

**The Model**

We model stabilizing selection in a multi-dimensional phenotype space, akin to Fisher's geometric model (71). An individual's phenotype is a vector in an *n*-dimensional Euclidian space, in which each dimension corresponds to a continuous quantitative trait. We focus on the architecture of one of these traits (say, the 1st dimension), where the total number of traits parameterizes pleiotropy. Fitness is assumed to decline with distance from the optimal phenotype positioned at the origin, thereby introducing stabilizing selection. Specifically, we assume that absolute fitness takes the form

$$W(\vec{r}) = \exp\left(-\frac{r^2}{2w^2}\right), \qquad (1)$$

where $\vec{r}$ is the (*n*-dimensional) phenotype, $r = \|\vec{r}\|$ is the distance from the origin and $w$ parameterizes the strength of stabilizing selection. However, we later show that the specific form of the fitness function doesn't matter. Moreover, the additive environmental contribution to the phenotype can be absorbed into $w$ (72; Section 1.1 in S1 Text); we therefore consider only the genetic contribution.

The genetic contribution to the phenotype follows from the multi-dimensional additive model (73). Specifically, we assume that the number of genomic sites affecting the phenotype (the target size) is very large, $L \gg 1$, and that allelic effects on the phenotype at these sites are vectors in the $n$-dimensional trait space. An individual's phenotype then follows from adding up the effects of her or his alleles, i.e.,

$$\vec{r} = \sum_{l=1}^{L}(\vec{a}_l + \vec{a}_l'), \qquad (2)$$

where $\vec{a}_l$ and $\vec{a}_l'$ are the phenotypic effects of the parents' alleles at site $l$.

The population dynamics follows from the standard model of a diploid, panmictic population of constant size $N$, with non-overlapping generations. In each generation, parents are randomly chosen to reproduce with probabilities proportional to their fitness (i.e., Wright-Fisher sampling with viability selection), followed by mutation, free recombination (i.e., no linkage) and Mendelian segregation. We further assume that the mutation rate per site, u, and the population size are sufficiently small that no more than two alleles segregate at any time at each site (i.e., that $\theta = 4Nu \ll 1$) and therefore an infinite sites approximation applies. The number of mutations per gamete, per generation therefore follows a Poisson distribution with mean $U = Lu$; based on biological considerations (see Sections 4.1 and 4.2 in S1 Text), we also assume that $1 \gg U \gg 1/2N$. The size of mutations in the $n$-dimensional trait space, $a\ (= \|\vec{a}\|)$, is drawn from some distribution, assuming only that $a^2 \ll w^2$. We later show that this requirement is equivalent to the standard assumption about selection coefficients satisfying $s \ll 1$ (also see Section 4.3 in S1 Text). The directions of mutations are assumed to be isotropic, i.e., uniformly distributed on the hypersphere in $n$-dimensions defined by their size, although we later show that our results are robust to relaxing this assumption as well.

## Results

**The phenotypic distribution.** In the first three sections, we develop the tools that we later use to study genetic architecture. We start by considering the equilibrium distribution of phenotypes in the population and generalizing previous results for the case with a single trait (26, 65, 66, 69). Under biologically sensible conditions, this distribution is well approximated by a tight multivariate normal centered at the optimum. Namely, the distribution of *n*-dimensional phenotypes, $\vec{r}$, in the population, is well approximated by the probability density function:

$$f(\vec{r}) = \frac{1}{(2\pi\sigma^2)^{n/2}} \exp\left(-\frac{r^2}{2\sigma^2}\right), \quad (3)$$

where $\sigma^2$ is the genetic variance of the phenotypic distances from the optimum (see Eq. S25 for closed form); and under plausible assumptions about the rate and size of mutations (i.e., when $1 \gg U \gg 1/2N$ and $a^2 \ll w^2$), it satisfies $\sigma^2 \ll w^2$, implying small variance in fitness in the population (Section 4.2 in S1 Text). Intuitively, the phenotypic distribution is normal because it derives from additive and (approximately) independently and identically distributed contributions from many segregating sites. Moreover, the population mean remains extremely close to the optimum because stabilizing selection becomes increasingly stronger with the displacement from it, and because any displacement is rapidly offset by minor changes to allele frequencies at many segregating sites.

With phenotypes close to the optimum, only the curvature of the fitness function at the optimum (i.e., the multi-dimensional second derivative) affects the selection acting on individuals. In addition, it is always possible to choose an orthonormal coordinate system centered at the optimum, in which the trait under consideration varies along the first coordinate and a unit change in other traits (along other coordinates) near the optimum have the same effect on fitness. These considerations suggest that the equilibrium behavior is insensitive to our choice of fitness function around the optimum. Moreover, in S1 Text (Section 5), we show that the rapid offset of perturbations of the population mean from the optimum (by minor changes to allele frequencies at numerous sites) lends robustness to the equilibrium dynamics with respect to the presence of major loci, moderate changes in the optimal phenotype over time and moderate asymmetries in the mutational distribution.

**Allelic dynamic.** Next, we consider the dynamic at a segregating site, and generalize previous results for the case with a single trait (67-69). This dynamic can be described in terms of the first two moments of change in allele frequency in a single generation (see, e.g., (74)). To calculate these moments for an allele with phenotypic effect $\vec{a}$ and frequency $q$ (=1-$p$), we note that the phenotypic distribution can be well approximated as a sum of the expected contribution of the allele to the phenotype, $2q\vec{a}$, and the distribution of contributions to the phenotype from all other sites, $\vec{R}$. From Eq. 3, it then follows that the distribution of background contributions is well approximated by probability density:

$$f(\vec{R}|\vec{a},q) = \frac{1}{(2\pi\sigma^2)^{n/2}} \exp\left(-\frac{(\vec{R}+2q\vec{a})^2}{2\sigma^2}\right). \tag{4}$$

By averaging the fitness of the three genotypes at the focal site over the distribution of genetic backgrounds, we find that the first moment is well approximated by

$$E(\Delta q) \approx \frac{a^2}{w^2} pq\left(q - \frac{1}{2}\right), \tag{5}$$

assuming that $a^2$ and $\sigma^2 \ll w^2$ (Section 4 in S1 Text). By the same token, we find that

$$V(\Delta q) \approx \frac{pq}{2N}, \tag{6}$$

which is the standard second moment with genetic drift.

The functional form of the first moment is equivalent to that of the standard viability selection model with under-dominance. This result is a hallmark of stabilizing selection on (additive) quantitative traits: with the population mean at the optimum, the dynamics at different sites are decoupled and selection at a given site acts to reduce its contribution to the phenotypic variance ($2a^2pq$), thereby pushing rare alleles to loss. Comparison with the standard viability selection model shows that the selection coefficient in our model is $s=a^2/w^2$, or $S=2Ns=2Na^2/w^2$ in scaled units. In other words, the selection acting on an allele is proportional to its size-squared in the *n*-dimensional trait space (where *w* translates effect size into units of fitness).

**The relationship between selection and effect size.** The statistical relationship between the strength of selection acting on mutations and their effect on a given trait follows from the aforementioned geometric interpretation of selection. Specifically, all mutations with a given selection coefficient, s, lie on a hypersphere in *n*-dimensions with radius $a = 2w\sqrt{s}$, and any given mutation satisfies

$$s = \frac{1}{w^2}a^2 = \frac{1}{w^2}\sum_{i=1}^{n} a_i^2, \tag{7}$$

where $a_i$ is the allele's effect on the i-th trait (Fig. 1A). Our assumption that mutation is isotropic then implies that the probability density of mutations on the hypersphere is uniform.

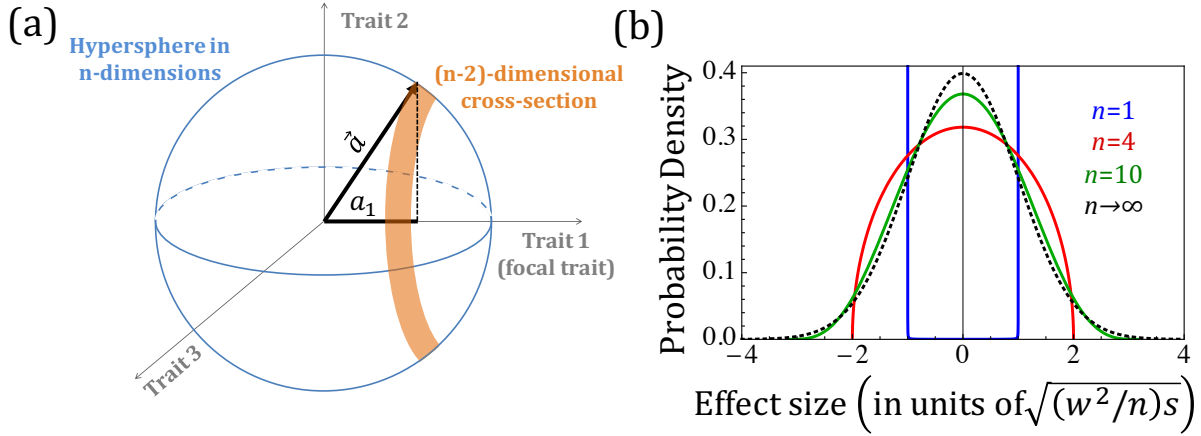

**Fig. 1.** The distribution of effect sizes corresponding to a given selection coefficient. (a) Mutations with selection coefficient, $s$, lie on a hypersphere in $n$ dimensions with radius $a = w\sqrt{s}$. The probability that such mutations have effect size $a_1$ on the focal trait is proportional to the volume of the $(n-2)$–dimensional cross section of the hypersphere, with projection $a_1$ on the coordinate corresponding to the trait. (b) The distribution of effect sizes on the focal trait, conditional on the selection coefficient being $s$, measured in units of the distribution's standard deviation (see Eq. 11).

The distribution of effect sizes on a focal trait, $a_1$, corresponding to a given selection coefficient, $s$, follows. Given that mutation is symmetric in any given trait, $E(a_1|s) = 0$, and given that it is symmetric among traits,

$$E(a_1^2|s) = a^2/n = (w^2/n)s. \tag{8}$$

More generally, the probability density corresponding to an effect size $a_1$ is proportional to the volume of the $(n-2)$ – dimensional cross section of the hypersphere with projection $a_1$ (Fig. 1A). For a single trait, this implies that $a_1 = \pm a$ with probability ½, and for $n > 1$, it implies the probability density

$$\varphi_n(a_1|a) = \frac{\Gamma(n/2)/\Gamma((n-1)/2)}{\sqrt{n/2}} \frac{1}{\sqrt{2\pi(a^2/n)}} \left(1 - \frac{1}{n}\frac{a_1^2}{(a^2/n)}\right)^{\frac{n-3}{2}} \tag{9}$$

(Section 1.2 in S1 Text). Intriguingly, when the number of traits $n$ increases, this density approaches a normal distribution, i.e.,

$$\frac{a_1}{\sqrt{a^2/n}} \sim N(0,1), \tag{10}$$

implying that the distribution of effect sizes given the selection coefficient becomes

$$a_1 \sim N(0, (w^2/n)s). \tag{11}$$

This limit is already well approximated for a moderate number of traits (e.g., $n$=10; Fig. 1B).

The limit behavior also holds when we relax the assumption of isotropic mutation. This generalization is important because, having chosen a parameterization of traits in which the fitness function near the optimum is isotropic, we can no longer assume that the distribution of mutations is also isotropic (75). Specifically, mutations might tend to have larger effects on some traits than on others, and their effects on different traits might be correlated. In Section 5.4 in S1 Text, we show that the limit distribution (Eq. 11) also holds for anisotropic mutation (excluding pathological cases). To this end, we introduce the concept of an effective number of traits, $n_e$, which can take any real value ≥1, and is defined as the number of equivalent traits required to generate the same relationship between the strength of selection on mutations and their expected effects on the trait under consideration (i.e., replacing $n$ in Eq. 11). The robustness of our model, along with mounting evidence that genetic variation is highly pleiotropic (see Introduction), suggest that the limit form may apply quite generally. In that regard, we note that even in this limit, the strength of selection on mutations and their effects on the focal trait are correlated, implying that the kind of "purely pleiotropic" extreme postulated in previous work cannot arise (18-20).

**Genetic architecture.** We can now derive closed forms for summary statistics of the genetic architecture (see Section 2.3 in S1 Text). For mutations with a given selection coefficient, the frequency distribution follows from the diffusion approximation based on the first two moments of change in allele frequency (Eqs. 5 and 6; (74)), and the distribution of effect sizes follows from the geometric considerations of the previous section. Conditional on the selection coefficient, these distributions are independent and therefore the joint distribution of frequency and effect size equals their product. Summaries of architecture can be expressed as expectations over the joint distribution of frequencies and effect sizes for a given selection coefficient, and then weighted according to the distribution of selection coefficients. While we know little about the distribution of selection coefficients of mutations affecting quantitative traits, we can draw general conclusions from examining how summaries of architecture depend on the strength of selection.

**Expected variance per site.** We focus on the distribution of additive genetic variance among sites, a central feature of architecture that is key to connecting our model with GWAS results. We start by considering how selection affects the expected contribution of a site to additive genetic variance in a focal trait. We include monomorphic sites in the expectation, such that the expected total variance is given by the product of the expectation per-site and the population mutation rate, $2NU$. Under the infinite sites assumption, sites are monomorphic or bi-allelic and their expected contribution to variance is

$$\mathrm{E}(2a_1^2 pq|S) = \mathrm{E}(a_1^2|S)\mathrm{E}(2\,pq|S) = \frac{w^2}{2Nn} S\, \mathrm{E}(2pq|S) \qquad (12)$$

(expressed in terms of the scaled selection coefficient $S$). Thus, the degree of pleiotropy only affects the expectation through a multiplicative constant.

This multiplicative factor would have a discernable effect in generalizations of our model in which the degree of pleiotropy varies among sites. For example, if the degree of pleiotropy of one set of sites was $k$ and of another set was $l > k$, and both sets were subject to the same strength of selection, then the expected contribution to genetic variance of sites in the first set would be $l/k$ times greater than in the second (from Eq. 12). While such generalizations may prove interesting in the future, here we focus on the model in which the degree of pleiotropy is constant. In this case, the multiplicative factor introduced by pleiotropy is not identifiable from data, because even if we could measure genetic variance in units of fitness (e.g., rather than in units of the total phenotypic variance), we still would not be able to distinguish between the effects of $w$ and $n$ on the genetic variance per site. We therefore focus on the effect of selection on the relative contribution to variance, which is insensitive to the degree of pleiotropy in our model.

The effect of selection on the relative contribution to genetic variance was described by Keightley and Hill (in the one dimensional case (67)) and is depicted in Fig. 2A. When selection is strong (roughly corresponding to $S > 30$), its effect on allele frequency (which scales with $1/S$) is canceled out by its relationship with the effect size (Eq. 8), yielding a constant contribution to genetic variance per site, $v_S = 2w^2/nN$, regardless of the selection coefficient (Section 3.1 in S1 Text; Figs. 2A and S1B). Henceforth, we measure genetic variance in units of $v_S$. When selection is effectively neutral (roughly corresponding to $S < 1$) and thus too weak to affect allele frequency, the expected contribution of a site to genetic variance scales with the effect size and equals ½$S$ ($\cdot v_S$), and therefore is lower than under strong selection (Section 3.1 in S1 Text; Figs. 2A and S1A). In between these selection regimes, selection effects on allele frequency are more complex and are influenced by under-dominance (Section 3.1 in S1 Text). As the selection coefficient increases, the expected contribution to variance reaches $v_S$ at $S \approx 3$, and continues to increase until it reaches a maximal contribution that is approximately 30% greater at $S \approx 10$ (Fig. 2A), after which it slowly declines to the asymptotic value of $v_S$ (Figs. 2A and S1B). Henceforth, we refer to this selection regime as intermediate (not to be confused with the nearly neutral range, which is much narrower and does not include selection coefficients with $S > 10$). These results suggest that effectively neutral sites should contribute much less to genetic variance than intermediate and strongly selected ones (66, 67).

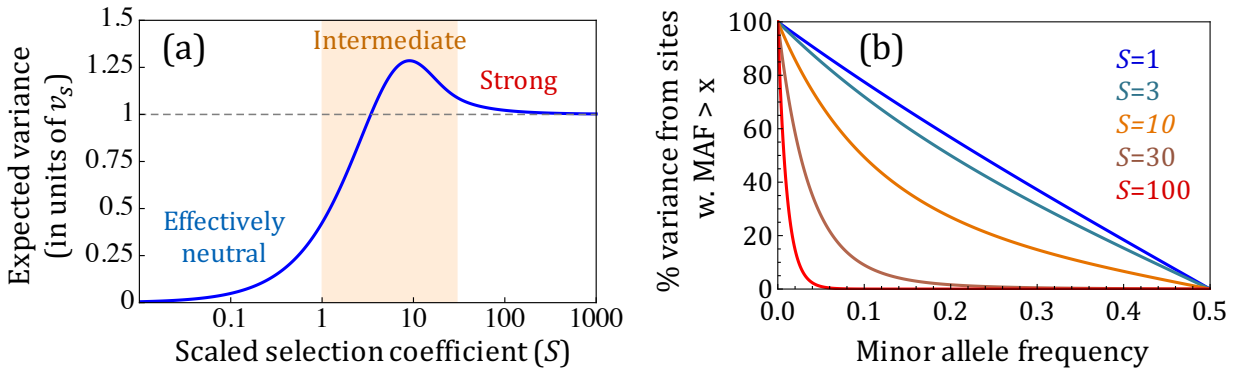

**Fig. 2.** The distribution of additive genetic variance among sites. In (a), we plot the expected contribution as a function of the scaled selection coefficient. We measure genetic variance in units of $v_S$ – the expected contribution at sites under strong selection. In (b), we show the proportion of additive genetic variance that arises from sites with MAF greater than the value on the x-axis, for different intermediate and strong selection coefficients.

While intermediate and strongly selected sites contribute similarly to variance, their minor allele frequencies (MAF) can differ markedly (Fig. 2B). As an illustration, segregating sites with MAF > 0.1 account for ~72% and ~49% of the additive genetic variance for intermediate selection coefficients of $S$=3 and 10, respectively, when almost no segregating sites would be found at such high MAF for a strong selection coefficient of $S$=100 (Fig. 2B). Thus, within the wide range of selection coefficients characterized as intermediate and strong, genetic variance arises from sites segregating at a wide range of MAF ranging from common to exceedingly rare.

**Distribution of variances among sites.** Next, we consider how genetic variance is distributed among sites with a given selection coefficient. We focus on the distribution among segregating sites (including monomorphic effects would just add a point mass at 0). This distribution is especially relevant to interpreting the results of GWAS, because, to a first approximation, a study will detect only sites with contributions to variance exceeding a certain threshold, $v = 2a_1^2 pq$, which decreases as the study size increases (see Discussion). We therefore depict the distribution in terms of the proportion of genetic variance, $G(v)$, arising from sites whose contribution to genetic variance exceeds a threshold $v$.

We begin with the case without pleiotropy ($n=1$), in which selection on an allele determines its effect size (Fig. 3A). When selection is strong ($S>30$), the proportion of genetic variance exceeding a threshold $v$ is also insensitive to the selection coefficient and takes a simple form, with

$$G(v) = \exp(-2v) \quad (13)$$

(Fig. 3A; Section 3.2 in S1 Text). In contrast, in the effectively neutral range ($S < 1$),

$$G(v) = \sqrt{1 - v/v_{max}}, \quad (14)$$

where the dependency on the selection coefficient enters through $v_{max} = \frac{1}{8}S$, which is the maximal contribution to variance and corresponds to an allele frequency of ½ (Fig. S4A; Section 3.2 in S1 Text). In the intermediate selection regime, $G(v)$ is also intermediate and takes a more elaborate functional form (Section 3.2 in S1 Text). These results suggest how genetic variance would be distributed among sites given any distribution of selection coefficients (Fig. 3A): starting from sites that contribute the most, the distribution would at first be dominated by strongly selected sites, then the intermediate selected sites would begin to contribute, whereas effectively neutral sites would enter only for $v < \frac{1}{8}S \ll 1$.

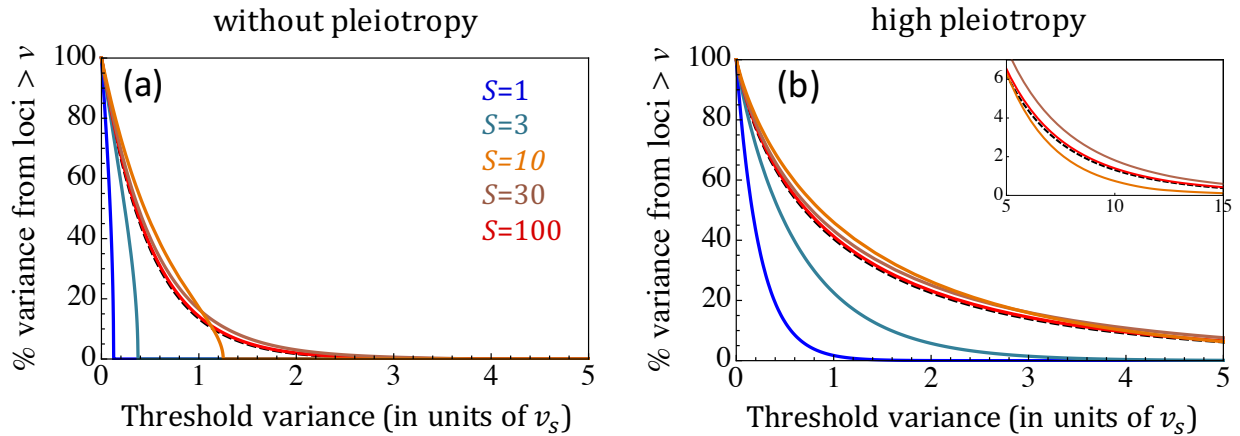

**Fig. 3.** The proportion of additive genetic variance that arises from sites that contribute more than the value on the x-axis, for a single trait (a) and in the pleiotropic limit (b). Our approximations for sites under strong selection (Eqs. 12 & 14) are shown with the dashed black curves. For the approximations in the effectively neutral limit (Eqs. 14 and 16), see Fig. S4.

Pleiotropy causes sites with a given selection coefficient to have a distribution of effect sizes on the focal trait, thereby increasing the contribution to genetic variance of some sites and decreasing it for others. In Section 3.2 of S1 Text, we show that increasing the degree of pleiotropy, $n$, increases the proportion of genetic variance, $G(v)$, for any threshold, $v$, regardless of the distribution of selection coefficients (Fig. S5). When variation in a trait is sufficiently pleiotropic for the distribution of effect sizes to attain the limit form (Eq. 11):

$$G(v) = \left(1 + 2\sqrt{v}\right)\exp\left(-2\sqrt{v}\right) \qquad (15)$$

for strongly selected sites and

$$G(v) = \exp(-4\,v/S) \qquad (16)$$

for effectively neutral ones (Figs. 3B and S4B; Section 3.2 in S1 Text). The intermediate selection range is split between these behaviors: on the weaker end, roughly corresponding to $S < 5$, $G(v)$ is similar to the effectively neutral case (Fig. S4B and Section 3.2 in S1 Text); and on the stronger end, roughly corresponding to $S > 5$, $G(v)$ is similar to the case of strong selection, with measurable differences only when $v \gg v_s$ (inset in Fig. 3B and Section 3.2 in S1 Text). We would therefore expect that as the sample size of GWAS increases and the threshold contribution to variance decreases, intermediate and strongly selected sites (more precisely sites with $S > 5$) will be discovered first, and effectively neutral sites will be discovered much later. In S1 Text (Section 3.2 and Fig. S3), we also derive corollaries for the distribution of numbers of segregating sites that make a given contribution to genetic variance.

## Discussion

**Interpreting the results of human GWAS.** In humans, GWAS for many traits display a similar behavior: when sample sizes are small, the studies discover almost nothing, but once they exceed a threshold sample size, both the number of associations discovered and the heritability explained begin to increase rapidly (4, 76). Intriguingly though, both the threshold study size and rate of increase vary among traits. These observations raise several questions, including: How is the threshold study size determined? How should the number of associations and explained heritability increase with study size once this threshold is exceeded? With an order of magnitude increase in study sizes into the millions imminent, how much more of the genetic variance in traits should we expect to explain? The theory that we developed provides tentative answers to these questions.

To relate the theory to GWAS, we must first account for the power to detect loci that contribute to quantitative genetic variation. In studies of continuous traits, the power can be approximated by a step function, where loci that contribute more than a threshold value $v^*$ to additive genetic variance will be detected and those that contribute less will not (Section 6.1 in S1 Text). The threshold depends on the study size, $m$, and on the total phenotypic variance in the trait, $V_P$, where $v^*(m, V_P) \propto V_P/m$ (Section 6.1 in S1 Text; 76). Given a trait and study size, the number of associations discovered and heritability explained then follow from our predictions for the distribution of variances among sites.

When genetic variation in a trait is sufficiently pleiotropic, our results suggest that the first loci to be discovered in GWAS will be intermediate or strongly selected, with correspondingly large effect sizes (i.e., $S \approx \frac{2Nn}{w^2} a_1^2 > 5$). The threshold study size for their discovery is proportional to $V_P/v_s$, i.e., the total phenotypic variance, $V_P$, measured in units of the expected contribution to variance of strongly selected sites, $v_s$ (Fig. 4). Beyond this study size, the number of associations detected and proportion of variance explained depend on the threshold variance $v^*$ measured in units of $V_P/v_s$, and follow from the functional forms that we derived for intermediate and strongly selected sites (Eq. 15 and Table S1). The dependence on $V_P/v_s$ makes intuitive sense, as the total phenotypic variance is background noise for the discovery of individual loci. Some results are modified when variation in a trait is only weakly pleiotropic, which is probably less common: notably, the threshold study size for strongly selected loci would be higher and loci under intermediate selection would begin to be discovered only after the strongly selected ones (Fig. S22 and Eqs. 13 and S35). Regardless of the degree of pleiotropy, effectively neutral loci would only begin to be discovered at much larger study sizes, after the bulk of intermediate and strongly selected variance has been mapped. Thus, the dependence of the explained heritability on study size is largely determined by $V_P/v_s$ and by the proportion of heritable variance arising from intermediate and strongly selected loci, whereas the number of associations also depends on the mutational target size, providing an explanation for why the performance of GWAS varies among traits.

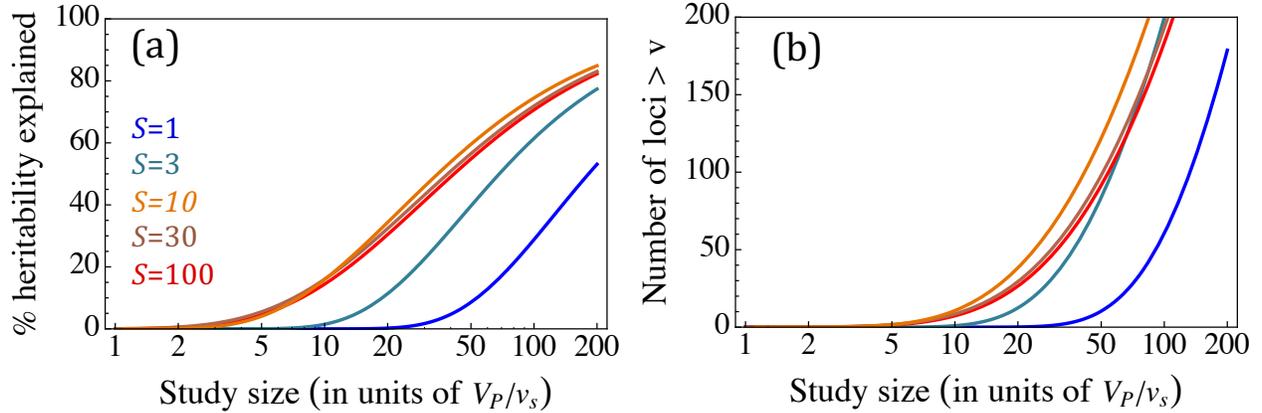

**Fig. 4.** The proportion of heritability (a) and the number of variants per Mbp (b) identified in GWAS as a function of study size, in the pleiotropic limit (see Sections 3.3 and 6.1 in S1 Text for derivations). For the case without pleiotropy, see Fig. S22.

**Inference and prediction.** Importantly, these theoretical predictions can be tested. As an illustration, we consider height and body mass index (BMI) in Europeans, two traits for which GWAS have discovered a sufficiently large number of genome-wide significant (GWS) associations (697 for height (16) and 97 for BMI (15)) for our test to be well powered. We fit our theoretical predictions to the distributions of variances among GWS associations reported for each of these traits, assuming that these distributions faithfully reflect what they would look like for the true causal loci (see Section 6.3 in S1 Text). We further assume that these loci are under intermediate or strong selection (as our predictions suggest) and that they are highly pleiotropic (see Introduction; 36, 41, 44). Under these assumptions, we expect the distribution of variances to be well approximated by a simple form (Eq. S89), which depends on a single parameter, $v_s$. We find that the theoretical distribution with the estimated $v_s$ fits the data for both traits well (Fig. 5A): we cannot reject our model based on the data for either trait (by a Kolmogorov-Smirnov test, $p = 0.14$ for height and $p = 0.54$ for BMI; Section 7.5 in S1 Text). By comparison, without pleiotropy ($n=1$), our predictions provide a poor fit to these data (by a Kolmogorov-Smirnov test, $p < 10^{-5}$ for height and $p = 0.05$ for BMI; Fig. S14).

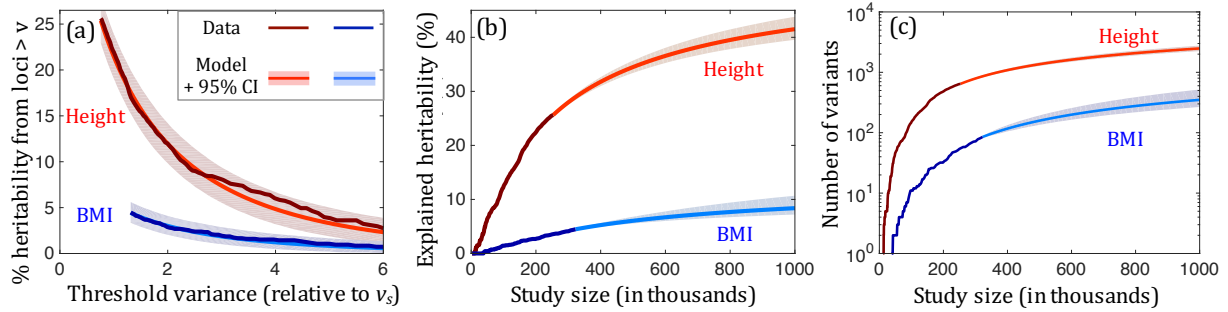

**Fig. 5.** Model fit and predictions for height and BMI, based on data from (16) and (15), respectively. In (a), we show the fit for associated loci. In (b) and (c), we show our predictions for future increases in the heritability explained and number of variants identified as GWAS size increases. 95% CIs are based on bootstrap; see Section S7.4 in S1 Text for details.

Fitting the model to GWAS results further allows us to make inferences about evolutionary parameters (Sections 7.1 and 7.3 in S1 Text). By including the degree of pleiotropy ($n$) as an additional parameter, we find that for both height and BMI, $n$ is sufficiently large for it to be indistinguishable from the high pleiotropy limit. Based on the shape of the fitted distributions in this limit and on the threshold values of $v^*$ (in units of $v_s$), we estimate that the proportion of variance arising from mutations within the range of detectable selection effects is ~50% for height and ~15% for BMI. Further relying on the number of associations that fall above the thresholds, we infer that, within this range, height has a mutational target size of ~5 Mb, whereas BMI has a target size of ~1 Mb (Table S2 in S1 Text).

These parameter estimates can help to interpret GWAS results. They suggest that, despite their comparable sample sizes, the GWAS for height succeeded in mapping a substantially greater proportion of the heritable variance than the GWAS for BMI (~20% compared to ~3-5%) because the proportion of variance arising from mutations within the range of detectable selection effects for height is much greater than for BMI. Moreover, the estimates of target sizes and the relationship between sample size and threshold contribution to variance can be used to predict how the explained heritability and number of associations should increase with sample size (Fig. 5B-C). These predictions are likely under-estimates as the range of detectable selection effects itself should also increase with sample size.

We can also examine to what extent our inferences are consistent with data and estimates from earlier studies. For example, the distribution of variances that we inferred for height fits those obtained in a recent GWAS of height based on exome genotyping (Kolmogorov-Smirnov test, $p = 0.99$; Fig. S15B and Section 8.1 in S1 Text). In addition, the proportion of variance that we estimate to arise from the range of selection effects detectable in existing GWAS for height and BMI are consistent with estimates of the heritable variance tagged by all SNPs with MAF>1% (59, 60; Section 8.2 in S1 Text).

**The effect of polygenic adaptation.** While we have assumed that quantitative traits have been subject to long-term stabilizing selection, recent studies indicate that some traits, and height in particular, have also been subject to recent directional selection (77-81). Under plausible evolutionary scenarios, recent directional selection can induce large changes to the mean phenotype through the collective response at many segregating loci, while having a negligible effect on allele frequencies at individual loci (21, 82). This very subtle effect on allele frequencies is likely one reason why polygenic adaptation is so difficult to detect, and why studies have to pool faint signals across many loci to do so (77-81). In Section 5.1 of S1 Text, we show that the distribution of allele frequencies on which our results rely is insensitive to sizable recent changes to the optimal phenotype. Importantly then, even when recent directional selection has occurred and its effects are discernable, the genetic architecture of a trait is nonetheless likely to be dominated by the effects of longer-term stabilizing selection.

**The effect of demography.** In contrast, recent changes in the effective population size are likely to have had a dramatic effect on allele frequencies and thus on the genetic architecture of quantitative traits (83, 84). In particular, European populations in which the GWAS for height and BMI were performed are known to have experienced dramatic changes in population size, including an Out of Africa (OoA) bottleneck ~100 KYA and explosive growth over the past ~5 KY (85-88). To study how these changes would have affected genetic architecture, we simulated allelic trajectories under our model and historical changes in population sizes in Europeans (relying on the model of (88); Section 9 in S1 Text).

Our results suggest that the individual segregating sites with the greatest contribution to the genetic variance that exists at present have selection coefficients around $s = 10^{-3}$ and are due to mutations that originated shortly before or during the OoA bottleneck (Fig. 6A and Section 9 in S1 Text). These mutations ascended to relatively high frequencies during the bottleneck and minimally decreased in frequency during subsequent, recent increases in population size, thereby resulting in large contributions to current genetic variance. Segregating sites under weaker selection contribute much less to variance because of their smaller effect sizes (i.e., for the same reason that applied in the case with a constant population size). Finally, and in contrast to the case with a constant population size, individual segregating sites under stronger selection (e.g., $s \geq 10^{-2.5}$) contribute much less to current variance than those with $s \approx 10^{-3}$. Mutations at these sites arose since the bottleneck, when the population size was considerably larger, resulting in much lower initial and current frequencies, and therefore a lower per (segregating) site contribution to variance (as distinct from the proportion of strongly selected sites that are currently segregating, which will have greatly increased, resulting in the same total contribution to variance; 83, 84). In Section 10 in S1 Text, we discuss one implication of this result: that the reliance on genotyping rather than resequencing in GWAS had practically no effect on the discovery of associations.

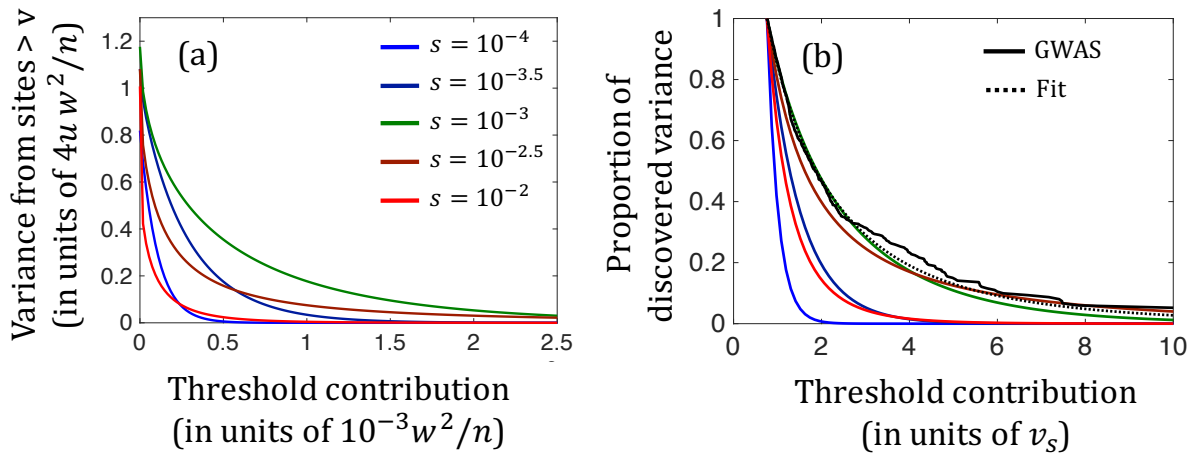

**Fig. 6.** The combined effect of selection and changes in population size (as inferred by (88) for Europeans) on the distribution of variances among segregating sites. (a) The cumulative variance arising from sites with contributions above a threshold as a function of the threshold, for different selection coefficients. Cumulative variance is measured in units of $4u \cdot w^2/n$, the equilibrium expectation for a strongly selected site, while the threshold is in units of $10^{-3} \cdot w^2/n$. (b) The distribution of variances among loci identified in the GWAS of height. The empirical distribution is in solid black and our inferred fit is in dashed black. Simulation results for each selection coefficient (in color) are normalized such that the proportion of variance at the study threshold is always 1. For similar results corresponding to BMI, see Fig. S20B, and for further details see Section 9 in S1 Text.

Segregating loci with $s \approx 10^{-3}$ not only make the largest contributions to the current variance, but are also likely to account for most of the GWS associations in the GWAS of height and BMI (Section 9 in S1 Text). When we account for the discovery thresholds of these studies, the expected distribution of variances for loci with $s \approx 10^{-3}$ closely matches the distribution observed among GWS associations (Figs. 6B & S20B). Moreover, these distributions closely match our theoretical predictions for $s \approx 10^{-3}$ and an $N_e \approx 5000$ (Fig. 6B)—roughly the effective population size experienced by mutations that originated shortly before or during the bottleneck. This match likely explains why the results predicted on a constant population size fit the data well nonetheless. Our interpretation of GWAS findings is supported by other aspects of the data (Section 9 in S1 Text).

Our conclusions about the high degree of pleiotropy of genetic variation for height and BMI and the differences between these traits are likely robust to demographic effects, given how well our model fits the distributions of variances among loci, once we account for European demographic history. However, we might be underestimating the mutational target sizes and total heritable variances associated with the selection effects currently visible in GWAS, as simulations with European demographic history indicate that the proportion of variance arising from loci with $s \approx 10^{-3}$ explained by current GWAS is lower than our equilibrium estimates (~42% compared to ~53% for height, and ~29% compared to ~38% for BMI). By the same token, we likely underestimated the future increase in explained heritability with increases in study sizes (Fig. 5B-C).

**Conclusion.** In summary, a ground-up model of stabilizing selection and pleiotropy can go a long way toward explaining the findings emerging from GWAS. Important next steps involve explicitly using more information from GWAS in the inferences. In particular, we can learn more about the selection acting on quantitative genetic variation by explicitly incorporating information about frequency and effect size (rather than their combination in terms of variance), and by including information from associations that do not attain genome-wide significance. Doing so will further require directly incorporating the effects of recent demographic history on genetic architecture (83, 84). An extended version of the inference, applied to the myriad traits now subject to GWAS, should allow us to learn about differences in the genetic architectures of traits, and answer long-standing questions about the evolutionary forces that shape quantitative genetic variation.

**Acknowledgements.** We have benefited hugely from discussions and comments from Guy Amster, Nick Barton, Jeremy Berg, Graham Coop, Laura Hayward, David Murphy, Joe Pickrell, Jonathan Pritchard and Molly Przeworski, as well as from helpful comments from four anonymous reviews. The work was funded by NIH grant GM115889 to GS.

**Software:** A Mathematica notebook calculating the main functions and reproducing the main figures is available at https://github.com/sellalab/GenArchitecture.

# A population genetic interpretation of GWAS findings

# for human quantitative traits

Supplementary Information

**Yuval B. Simons[a,1], Kevin Bullaughey[b], Richard R. Hudson[b], and Guy Sella[a,1]**

[a] Department of Biological Sciences, Columbia University, New York, NY 10027;

[b] Department of Ecology & Evolution, University of Chicago, Chicago, Illinois 60637;

[1] To whom correspondence should be addressed: yuval.simons@columbia.edu, gs2747@columbia.edu

## Table of Contents

## 1. The model

### 1.1. Absorbing the environmental contribution into the fitness function

Here, we show that the additive environmental contribution to the phenotype can be absorbed into the fitness function, which justifies our considering only the additive genetic contribution in our analysis. This result has been derived multiple times for the one dimensional case (e.g., 1). The argument in the multi-dimensional case is similar and included for completeness.

First, assume that the additive environmental contribution to the phenotype, $\vec{r}_e$, is distributed as a multi-normal with mean 0 and isotropic variance $V_e$. The expected absolute fitness of an individual with additive genetic contribution to the phenotype, $\vec{r}_g$, is given by averaging fitness over the distribution of environmental contributions. Namely,

$$\overline{W}(\vec{r}_g) = \int_{\vec{r}_e} \frac{1}{(2\pi V_e)^{n/2}} e^{-\frac{|\vec{r}_e|^2}{2V_e}} W(\vec{r}_g + \vec{r}_e) = \int_{\vec{r}_e} \frac{1}{(2\pi V_e)^{n/2}} e^{-\frac{|\vec{r}_e|^2}{2V_e}} e^{-\frac{|\vec{r}_g + \vec{r}_e|^2}{2w^2}}$$

$$= \frac{1}{(1+V_e/w^2)^{n/2}} e^{-\frac{|\vec{r}_g|^2}{2(w^2+V_e)}}. \tag{S1}$$

Given that absolute fitness is defined up to a multiplicative constant, we can therefore absorb the additive environmental contribution by using the Gaussian fitness function

$$\widetilde{W}(\vec{r}_g) = \exp\left(-\frac{|\vec{r}_g|^2}{2\widetilde{w}^2}\right), \tag{S2}$$

where $\widetilde{w}^2 = w^2 + V_e$. Even when the environmental contribution is anisotropic, we can always choose a coordinate system in which the effective fitness function takes an isotropic form around the fitness peak (Eq. 1).

### 1.2. The distribution of mutational effect sizes on a given trait

In the main text, we define the distribution of phenotypic effects of newly arising mutations in the $n$-dimensional trait space, $\vec{a}$. Here, we consider the projection of these effects on a given trait, $a_1$, taken without loss of generality to be on the 1st dimension. The distribution of effect sizes on a focal trait will depend on the degree of pleiotropy, $n$, and the form of this dependency becomes important when we consider how pleiotropy affects genetic architecture.

We want to calculate the distribution of effect sizes on the focal trait, $a_1$, conditional on their overall effect, $a = \|\vec{a}\|$. We assume that the distribution of effects of de novo mutations is isotropic in trait space. The effect of a mutation, $\vec{a}$, therefore has equal probability to occupy any point on an $n$-dimensional sphere with radius $a$. Let $S_m(x)$ denote the surface area of an $m$-dimensional sphere of radius $x$ and $\theta$ denote the angle between the vector $\vec{a}$ and its projection $a_1$, i.e., $a_1 = a \cos \theta$. In these terms, the surface area element corresponding to angle $\theta$ is

$$S_{n-1}(a \sin \theta) \, a \, d\theta, \tag{S3}$$

and by a change of variables, the surface area element corresponding to projection $a_1$ on the focal trait is

$$S_{n-1}(a \sin \theta) \, a \, d\theta = S_{n-1}\left(\sqrt{a^2 - a_1^2}\right) \frac{a}{\sqrt{a^2 - a_1^2}} da_1, \tag{S4}$$

since $da_1 = \left|\frac{da_1}{d\theta}\right| d\theta = a \sin \theta \, d\theta = \sqrt{a^2 - a_1^2} d\theta$. This result implies that the probability density of $a_1$ is

$$\varphi_n(a_1|a) = \frac{S_{n-1}\left(\sqrt{a^2 - a_1^2}\right)}{S_n(a)} \frac{a}{\sqrt{a^2 - a_1^2}} = \frac{\Gamma\left(\frac{n}{2}\right)}{\sqrt{\pi} \, \Gamma\left(\frac{n-1}{2}\right)} \left(1 - \frac{a_1^2}{a^2}\right)^{\frac{n-3}{2}} \frac{1}{a} \tag{S5}$$

(for a similar derivation, see (2)).

Next, we consider the high pleiotropy limit form of this distribution. For any degree of pleiotropy, the symmetry of the mutational distribution implies that

$$E(a_1|a) = 0 \tag{S6}$$

and the equivalence among traits implies that

$$V(a_1|a) = a^2/n \tag{S7}$$

(see main text for more details). It follows that when $n$ becomes sufficiently large, $a_1/a \ll 1$ and therefore

$$\left(1 - \frac{a_1^2}{a^2}\right)^{\frac{n-3}{2}} \approx \exp\left(-\frac{n}{2} \frac{a_1^2}{a^2}\right). \tag{S8}$$

In addition, $\Gamma\left(\frac{n}{2}\right)/\Gamma\left(\frac{n-1}{2}\right) \approx \sqrt{n/2}$. Substituting these expressions into Eq. S5, we find that for sufficiently large $n$ the distribution of effect sizes approaches the normal distribution

$$\varphi_n(a_1|a) \approx \frac{1}{\sqrt{2\pi(a^2/n)}} \exp\left(-\frac{1}{2} \frac{a_1^2}{a^2/n}\right). \tag{S9}$$

As we elaborate in the main text, important implications about quantitative genetic variation follow from this high pleiotropy limit. The limit also holds quite generally when the distribution of effect sizes is anisotropic (see Section S5.4).

## 2. Solving for summaries of genetic architecture

Here, we derive closed forms for summaries of genetic architecture under our model. We begin by deriving the first two moments of change in allele frequency in a single generation. With these moments at hand, we use the diffusion approximation to calculate the sojourn time for alleles that contribute to quantitative genetic variation (3). Together with the distribution of effect sizes derived in the previous section, the sojourn time allows us to obtain closed forms for summaries of genetic architecture. Specifically, we can obtain a closed form for any summary that can be described as a function of allele frequencies and effect sizes at sites contributing to quantitative genetic variation. We use these expressions to calculate the summaries used in the main text, for example the expected additive genetic variance and its distribution across sites.

### 2.1. The first two moments of change in allele frequency

We assume that:

- The phenotypic distribution at steady state is well approximated by an isotropic multivariate normal distribution centered at the optimum, namely by the probability density

$$f(\vec{r}) = \frac{1}{(2\pi\sigma^2)^{n/2}} \exp\left(-\frac{r^2}{2\sigma^2}\right). \tag{S10}$$

- Both $a^2$ and $\sigma^2 \ll w^2$.

These assumptions are justified in Section S3.3.

We rely on these assumptions to calculate the first two moments of change in frequency in a single generation for an allele with phenotypic effect $\vec{a}$ and frequency $q$. The fitnesses of the three genotypes at the site depend on its distribution of genetic backgrounds, i.e., on the total phenotypic contribution of sites other than the focal one $\vec{R}$. Following Eq. S10 and assuming every allele contributes only a small proportion of the genetic variance, the distribution of $\vec{R}$ is well approximated by

$$f(\vec{R}|\vec{a}, q) = \frac{1}{(2\pi\sigma^2)^{n/2}} \exp\left(-\frac{|\vec{R}+2q\vec{a}|^2}{2\sigma^2}\right). \tag{S11}$$

The expected fitnesses of the three genotypes then follow from integrating over backgrounds:

$$W_{00} = \int_{\vec{R}} f(\vec{R}|\vec{a}, q)W(\vec{R}) = \left(\frac{w}{\sqrt{w^2+\sigma^2}}\right)^n \exp\left(-\frac{4a^2q^2}{2(w^2+\sigma^2)}\right), \tag{S12}$$

$$W_{01} = \int_{\vec{R}} f(\vec{R}|\vec{a},q) W(\vec{R}+\vec{a}) = \left(\frac{w}{\sqrt{w^2+\sigma^2}}\right)^n \exp\left(-\frac{4a^2(q-1/2)^2}{2(w^2+\sigma^2)}\right) \quad (S13)$$

and

$$W_{11} = \int_{\vec{R}} f(\vec{R}|\vec{a},q) W(\vec{R}+2\vec{a}) = \left(\frac{w}{\sqrt{w^2+\sigma^2}}\right)^n \exp\left(-\frac{4a^2(q-1)^2}{2(w^2+\sigma^2)}\right). \quad (S14)$$

The first moment of change in allele frequency is then

$$E(\Delta q) = -pq \frac{p(W_{00}-W_{01})+q(W_{01}-W_{11})}{\overline{W}} \approx -\frac{a^2}{w^2} pq \left(\frac{1}{2}-q\right), \quad (S15)$$

relying on our assumptions that $a^2$ and $\sigma^2 \ll w^2$. The functional form of the first moment is equivalent to that of the standard viability selection model with under-dominance and selection coefficient $s = \frac{a^2}{w^2}$ or scaled selection coefficient

$$S = 2N\frac{a^2}{w^2}. \quad (S16)$$

Similarly, we find that

$$V(\Delta q) \approx \frac{pq}{2N}, \quad (S17)$$

which is the standard second moment with genetic drift.

## 2.2. Sojourn time

Based on the first two moments, we can use the diffusion approximation to calculate the sojourn time as a function of allele frequency, i.e., the density of the time that an allele spends at a given frequency $q$ before it fixes or is lost (3). For a mutant allele with initial frequency $1/2N$ and scaled selection coefficient $S$, the sojourn time is

$$\tau(q|S) = \begin{cases} \frac{N\sqrt{\pi/S}}{\text{erf}(\sqrt{S}/2)} \frac{e^{S(1-2x)^2/4}}{x(1-x)} f_-(S,q) f_+(S,1/2N) & 0 \leq q \leq 1/2N \\ \frac{N\sqrt{\pi/S}}{\text{erf}(\sqrt{S}/2)} \frac{e^{S(1-2x)^2/4}}{x(1-x)} f_+(S,q) f_-(S,1/2N) & 1/2N \leq q \leq 1 \end{cases} \quad (S18)$$

where erf is the error function and $f_\pm(S,y) \equiv \text{erf}(\sqrt{S}/2) \pm \text{erf}(\sqrt{S}(1-2y)/2)$.

The sojourn time takes simple limiting forms when selection is effectively neutral ($S \ll 1$) or strong ($S \gg 1$). In the effectively neutral range, it is well approximated by $\tau(q|S) = 2/q$, and in the strongly selected range, it is well approximated by $\tau(q|S) = 2\exp(-Sq)/q$.

## 2.3. Calculating expectations of summaries of architecture

Many summaries of interest can be expressed as sums over segregating sites of some function $c(q, a_1)$, where $q$ is the derived allele frequency and $a_1$ is the effect size on the trait. For example, the additive genetic variance in a trait is given by the sum of $v(q, a_1) = 2a_1^2 q(1-q)$ over sites. The expectation over such summaries can be expressed as

$$\mathrm{E}(C) = 2NU \int_q \int_{a_1} c(q, a_1) \rho(q, a_1), \tag{S19}$$

where $C$ is the summery summed over all sites, $2NU$ is the population mutation rate per generation and $\rho(q, a_1)$ is the density of sites with the corresponding frequency and effect size per unit mutational input.

The density $\rho(q, a_1)$ can be broken down into contributions from sites with different selection coefficients, i.e.,

$$\rho(q, a_1) = \int_S f(S)\bigl(\tau(q|S)\eta(a_1|S)\bigr), \tag{S20}$$

where $f(S)$ is the distribution of selection coefficients and $\tau(q|S)$ is the sojourn time of a mutation with selection coefficient $S$ (Eq. S18). The probability density $\eta(a_1|S)$ of effect sizes given selection coefficient $S$ follows from Eqs. S5 and S16

$$\eta(a_1|S) = \varphi_n(a_1|a(S)) = \varphi_n\left(a_1 \middle| \sqrt{(w^2/2N)S}\right). \tag{S21}$$

This allows us to break down our summaries into contributions from sites with different selection coefficients

$$\mathrm{E}(C) = 2NU \int_S f(S) \mathrm{E}(C|S) \tag{S22}$$

with

$$\mathrm{E}(C|S) = \int_q \int_{a_1} c(q, a_1) \tau(q|S) \eta(a_1|S). \tag{S23}$$

We use Eq. S23 to study how summaries of architecture depend on the strengths of selection, and how these summaries will depend on different distributions of selection coefficients. This allows us to draw general implications about genetic architecture despite our limited knowledge about this distribution.

## 3. Additive genetic variance and number of segregating sites

The distributions of additive genetic variance and of the number of segregating sites are critical to understanding genetic architecture and specifically to interpreting results of GWAS. Here we derive closed forms for both distributions as well as simple approximations under strong and effectively neutral selection.

### 3.1. Expectations

We begin by considering the expected contribution of a site to additive genetic variance. Substituting the contribution to variance from a single site $v(q, a_1) = 2a_1^2 q(1-q)$ into Eq. S23, we find that

$$E(V|S) = \int_q \int_{a_1} v(q, a_1)\tau(q|S)\eta(a_1|S) = \int_q 2q(1-q)\tau(q|S) \int_{a_1} a_1^2 \eta(a_1|S)$$

$$= \int_q 2q(1-q)\tau(q|S)\frac{w^2}{2Nn}S = \frac{2w^2}{nN}\int_q \frac{1}{2}Sq(1-q)\tau(q|S). \tag{S24}$$

The total additive genetic variance is

$$\sigma^2 = 2NU \int_S f(S)E(V|S). \tag{S25}$$

The closed form for $E(V|S)$ in Eq. S24 was integrated numerically to obtain Fig. 2A in the main text. We can use the results of Keightley and Hill (4) to obtain an analytic approximation for $E(V|S)$:

$$E(V|S) = \frac{2w^2}{nN}\sqrt{\frac{\pi S}{4}}\,\text{erfi}(\sqrt{S}/4)\exp(-S/4) + O\left(\frac{1}{2N}\right), \tag{S26}$$

where erfi is the imaginary error function ($\text{erfi}(x) \equiv \text{erf}(ix)/i$).

In the effectively neutral and strong selection limits, we can use limit forms of the sojourn time to derive simple approximations for $E(V|S)$. In the effectively neutral limit, i.e., when $S \ll 1$, $\tau(q|S) \approx 2/q$ and therefore

$$E(V|S) \approx \frac{2w^2}{nN}\frac{S}{2}. \tag{S27}$$

In practice, this expression provides a decent approximation when $S < 1$ (Fig. S1A). In the strong selection limit, when $S \gg 1$, $\tau(q|S) \approx 2\exp(-Sq)/q$ and therefore

$$E(V|S) \approx \frac{2w^2}{nN}. \tag{S28}$$

In practice, this expression provides a decent approximation when $S > 30$ (Fig. S1B). The constant $2w^2/nN$, which recurs in our derivations (e.g., Eq. S24), thus has a simple interpretation: it is the expected contribution of strongly selected sites to additive genetic variance (per unit mutational input). We therefore denote it by $v_s$, and henceforth measure variance in these units.

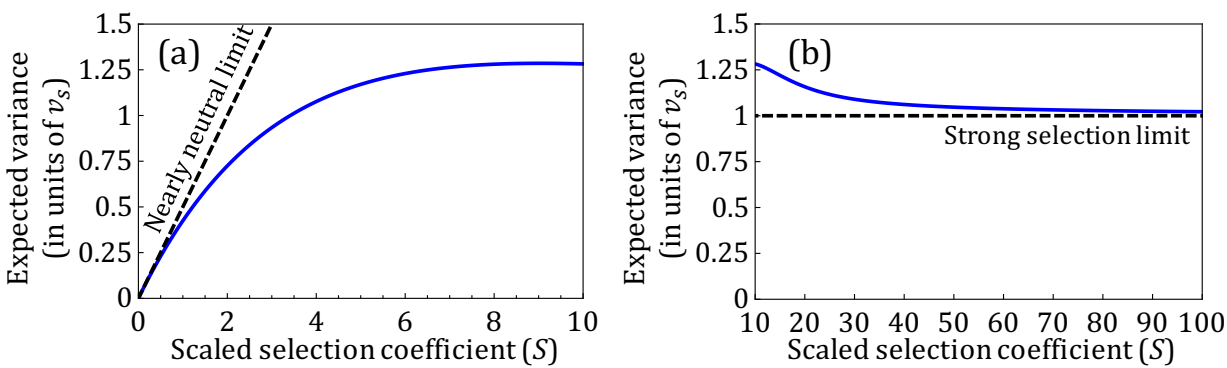

**Figure S1.** The effectively neutral and strong selection approximations for the expected contribution to genetic variance per site. (A) The expression in the limit of $S \ll 1$ provides a decent approximation when $S < 1$. (B) The expression in the limit of $S \gg 1$ provides a decent approximation when $S > 30$.

We next consider how the expected number of segregating sites depends on the strength of selection. This expectation (per unit mutational input) is simply the mean sojourn time of a newly arising mutation. Formally, it follows from substituting $c(q, a_1) = 1$ into Eq. S23, i.e.,

$$E(K|S) = \int_q \int_{a_1} \tau(q|S)\eta(a_1|S) = \int_q \tau(q|S) \int_{a_1} \eta(a_1|S) = \int_q \tau(q|S). \qquad (S29)$$

In Fig. S2, we calculate this integral numerically for different values of $S$, to find that the number of segregating sites depends only weakly on $S$. Intuitively, this follows from the fact that the vast majority of mutations, be they effectively neutral, intermediate, or strongly selected, spend only a few generations in the population at low copy numbers before going extinct.

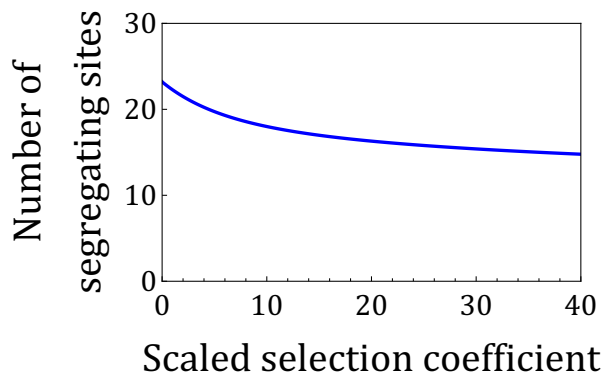

**Figure S2.** The number of segregating sites per unit mutational input (or, equivalently, the expected sojourn time of a newly arising mutation), Eq. S29, is only weakly dependent on the strength of selection. Calculated for a population size of 20,000.

## 3.2. Densities

Here, we consider how the additive genetic variance is distributed among sites. We begin by deriving a closed form for the density of segregating sites with a given contribution to variance $v$. This density follows from substituting Dirac's delta function $\delta(v - 2a_1^2 q(1-q))$ into Eq. S23:

$$\rho(v|S) = E\big(\delta(v - 2a_1^2 q(1-q))\big|S\big) = \int_q \int_{a_1} \delta(v - 2a_1^2 q(1-q))\tau(q|S)\eta(a_1|S) =$$

$$= \int_{a_1} \left(\tau(q_+(v,a_1)|S)\left|\frac{\partial q_+(v,a_1)}{\partial v}\right| + \tau(q_-(v,a_1)|S)\left|\frac{\partial q_-(v,a_1)}{\partial v}\right|\right)\eta(a_1|S)$$

$$= \int_{a_1} \big(\tau(q_+(v,a_1)|S) + \tau(q_-(v,a_1)|S)\big) \frac{1}{2a_1^2\sqrt{1-2v/a_1^2}} \eta(a_1|S), \tag{S30}$$

where $q_\pm(v,a_1) = \frac{1}{2}\left(1 \pm \sqrt{1 - 2v/a_1^2}\right)$ are the two frequencies for which $v = 2a_1^2 q(1-q)$. This integral can be calculated numerically for any $S$ and degree of pleiotropy $n$ (by using the corresponding density $\eta(a_1|S)$). Moreover, as we illustrate below, summary statistics of the distribution of variances among sites can be expressed and calculated in terms of integrals over the density $\rho(v|S)$.

We can greatly simplify the expression for $\rho(v|S)$ in the limits of effectively neutral and strong selection, and especially in the cases without pleiotropy or with extensive pleiotropy. When selection is effectively neutral ($S \ll 1$), then $\tau(q|S) \cong 2/q$ and thus

$$\rho(v|S) = \int_{a_1} \left(\frac{2}{q_+(v,a_1)} + \frac{2}{q_-(v,a_1)}\right)\frac{1}{2a_1^2\sqrt{1-2v/a_1^2}} \eta(a_1|S)$$

$$= \int_{a_1}\left(\frac{4}{1+\sqrt{1-2v/a_1^2}} + \frac{4}{1-\sqrt{1-2v/a_1^2}}\right)\frac{1}{2a_1^2\sqrt{1-2v/a_1^2}}\eta(a_1|S) = \int_{a_1}\frac{2}{v\sqrt{1-2v/a_1^2}}\eta(a_1|S), \tag{S31}$$

with variance measured in units of $v_s$ and effect size measured in units of $\sqrt{v_s}$. Without pleiotropy ($n = 1$), the effect size is $a_1 = \pm\frac{1}{2}\sqrt{S}$ and the expression for the density simplifies to

$$\rho(v|S) = \frac{2}{v\sqrt{1-v/v_{max}}}, \tag{S32}$$

where $v_{max} \equiv S/8$ is the maximal contribution to variance for a mutation with selection coefficient $S$, which is obtained when both alleles have frequency ½. When the degree of pleiotropy is high ($n \gg 1$), $a_1$ is approximately normally distributed with mean 0 and variance $S/4$ (Eq. 11) and the expression for the density simplifies to

$$\rho(v|S) = \int_{a_1 > \sqrt{2v}} \frac{2}{v\sqrt{1-2v/a_1^2}} \frac{2}{\sqrt{2\pi S/4}} \exp(-2a_1^2/S) = 2\exp(-4v/S)/v. \quad (S33)$$

When selection is strong, derived alleles are rare ($q \ll 1$), implying that $v \ll a_1^2$ and $q \approx v/2a_1^2$, and that the sojourn time is well approximated by $\tau(q|S) = 2\exp(-Sq)/q$. The density $\rho(v|S)$ then simplifies to

$$\rho(v|S) \approx \int_{a_1} \tau(q(v,a_1)|S) \frac{1}{2a_1^2} \eta(a_1|S) \approx \int_{a_1} \frac{4a_1^2}{v} \exp(-Sv/2a_1^2) \frac{1}{2a_1^2} \eta(a_1|S)$$

$$= \int_{a_1} \frac{2}{v} \exp(-Sv/2a_1^2) \eta(a_1|S). \quad (S34)$$

Without pleiotropy, this expression further simplifies to

$$\rho(v|S) \approx 2\exp(-2v)/v, \quad (S35)$$

and when the degree of pleiotropy is high ($n \gg 1$), then

$$\rho(v|S) \approx \int_{a_1} \frac{2}{v} \exp(-Sv/2a_1^2) \frac{2}{\sqrt{2\pi S/4}} \exp(-2a_1^2/S) = 2\exp(-2\sqrt{v})/v. \quad (S36)$$

We are especially interested in the distribution of variances among sites that exceed some threshold contribution $v^*$. As we discuss in the main text and in Section S6, to a first approximation, the loci identified in a GWAS would be those with contributions to additive variance that exceed the study's threshold contribution $v^*$. In particular, our inferences based on GWAS data rely on fitting the probability density of the number of segregating sites with variance $v$ that exceed a given threshold contribution $v^*$ (Section S7). This probability density is:

$$f(v|S) = \rho(v|S)/K(v^*|S), \quad (S37)$$

where

$$K(v^*|S) \equiv \int_{v > v^*} \rho(v|S) \quad (S38)$$

is the expected number of segregating sites with contributions to variance exceeding $v^*$ per unit mutational input.

In our analysis, we focus on the expected proportion of additive genetic variance arising from sites that exceed a threshold contribution $v^*$, which approximates the heritable variance explained in GWAS. This proportion is given by

$$G(v^*|S) = \frac{\int_{v>v^*} v\, \rho(v|S)}{\int_v v\, \rho(v|S)} = \frac{\int_{v>v^*} v\, \rho(v|S)}{E(V|S)}. \quad (S39)$$

Given a distribution of selection coefficients, $f(S)$, the corresponding proportion is

$$G(v^*) = \frac{\int_S G(v^*|S) E(V|S) f(S)}{\int_S E(V|S) f(S)} = \int_S G(v^*|S) \frac{E(V|S) f(S)}{\int_S E(V|S) f(S)}. \quad (S40)$$

The dependences of the proportion of variance $G(v^*|S)$ and the number of sites $K(v^*|S)$ on the strength of selection $S$ for cases without pleiotropy and with extensive pleiotropy are shown in Figs. 3 and S3, respectively. We rely on Eqs. S32, S33, S35, S36, S38 and S39 to derive simplified forms for these summaries in the effectively neutral and strongly selected limits (Table S1). While the expressions for the effectively neutral limit were derived for $S \ll 1$, in practice they provide a decent approximation when $S < 1$ (Fig. S4A & B). In the strong selection limit ($S \gg 1$), the expressions for the case without pleiotropy provide a decent approximation for $S > 30$ (Fig. 3A), whereas with extensive pleiotropy they already work quite well for $S > 5$ (Fig. 3B).

| Selection | Effectively neutral ($S \ll 1$) | | Strongly selected ($S \gg 1$) | |
|---|---|---|---|---|
| # of traits | $n = 1$ | $n \gg 1$ | $n = 1$ | $n \gg 1$ |
| $E(V|S)$ | $S/2$ | | 1 | |
| $G(v^*|S)$ | $\sqrt{1 - 8v^*/S}$ | $\exp(-4v^*/S)$ | $\exp(-2v^*)$ | $(1 + 2\sqrt{v^*})\exp(-2\sqrt{v^*})$ |
| $K(v^*|S)$ | $4 \cdot \text{artanh}(\sqrt{1 - 8v^*/S})$ | $2 \cdot I(4v^*/S)$ | $2 \cdot I(2v^*)$ | $4 \cdot I(2\sqrt{v^*})$ |

**Table S1.** Limits for the expected proportion of variance and expected number of sites corresponding to sites that exceed a threshold contribution to additive genetic variance $v^*$. $I(x) \equiv \int_{t>x} \exp(-t)/t$ is an exponential integral and artanh is the inverse hyperbolic tangent.

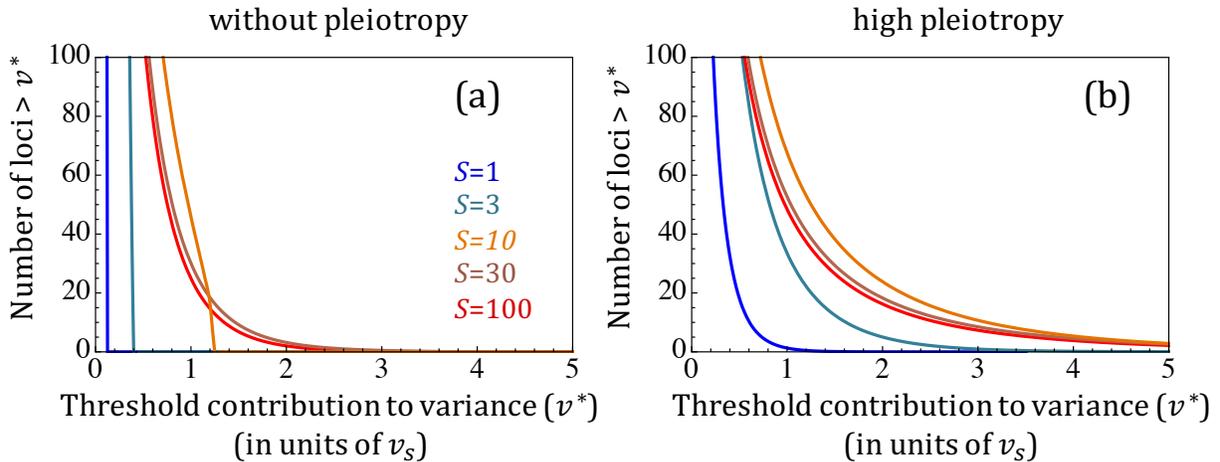

**Figure S3.** The number of loci per Mb contributing more than $v$ to the variance, as a function of $v$, (a) in the case without pleiotropy, $n = 1$, and (b) in the high pleiotropy limit, $n \gg 1$. We assume a constant population size of 20,000, with a mutation rate of $1.2 \cdot 10^{-8}$ (5).

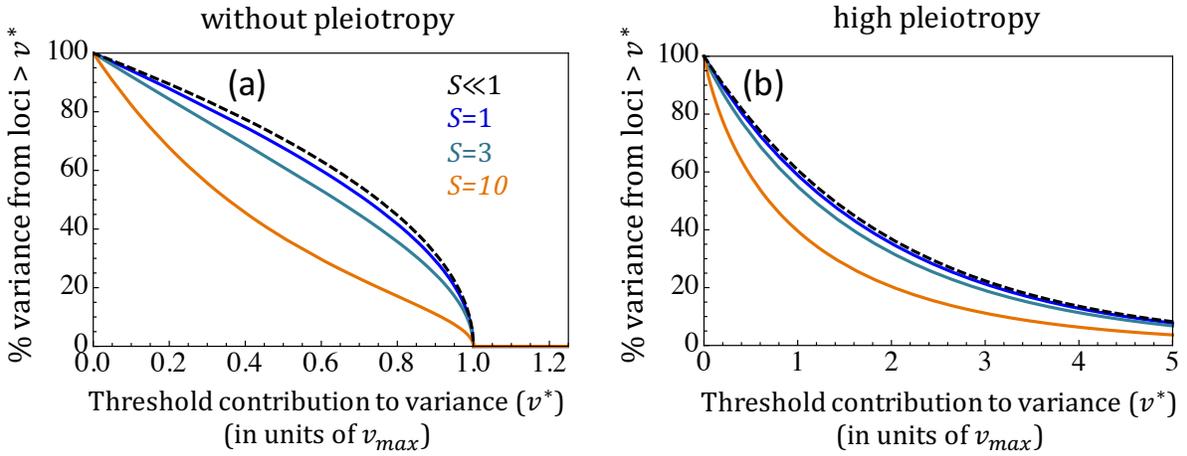

**Figure S4.** The proportion of additive genetic variance that arises from sites that contribute more than the value on the x-axis, for a single trait (a) and in the pleiotropic limit (b). We show the x-axis in units of $v_{max} = S/8$ (in units of $v_s$) in order to evaluate the approximations in the effectively neutral limit (in dashed black; Eqs. 14 & 16); note that $v_{max}$ is not the maximal variance in cases with pleiotropy.

Both the proportion of variance, $G(v^*|S)$, and number of variants, $K(v^*|S)$, appear to always increase with the degree of pleiotropy, $n$ (Fig. S5). We do not have a proof for this property but can suggest an intuitive explanation. Without pleiotropy ($n=1$), the selection coefficient determines the effect size, such that any contribution $v^*$ to genetic variance corresponds to a specific minor allele frequency $q^*$. The sites with contributions $v > v^*$ are therefore those with minor allele frequencies $q > q^*$. Pleiotropy causes sites with a given selection coefficient to have a distribution of effect sizes on the trait under consideration. As a result, some sites with frequencies above $q^*$ end up with contributions to variance below $v^*$ while others exceed $v^*$. To understand how this affects $G(v^*|S)$, recall that for any selection coefficient, the density of variants always rapidly increases as $q^*$ decreases. As long as the contribution $v^*$ and the corresponding frequency without pleiotropy $q^*$ are not close to 0, we may therefore expect that introducing pleiotropy would result in pushing more sites above $v^*$ than below $v^*$, resulting in a net increase to the proportion $G(v^*|S)$. For the same reasons, the number of variants with $v > v^*$ shows a similar behavior and also grows with $n$.

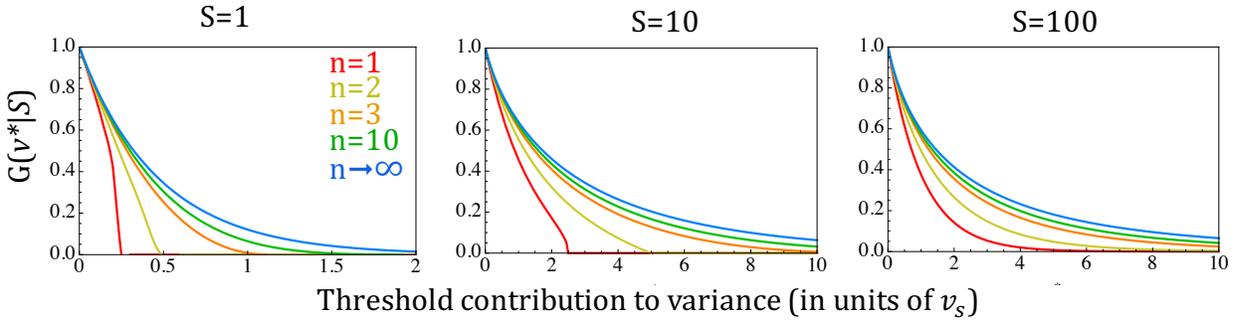

**Figure S5.** The effect of pleiotropy on the proportion of the variance explained by sites contributing more then $v^*$ to the variance, $G(v^*|S)$ (see Eq. S39). For all selection coefficients, the proportion of variance explained increases as the number of traits, i.e. the degree of pleiotropy, increases.

### 3.3. Comparing predictions against simulations

We tested our theoretical derivations for the total genetic variance and its distribution among sites against forward computer simulations. The code and documentation are available at https://github.com/sellalab/GenArchitecture. The simulation implements the model as specified in the main text, with a few additional details and one exception noted below. The additional details are that we assume the infinite sites model for mutation. Second, the distribution of scaled selection coefficients, or equivalently the distribution of mutation sizes (see Eq. 7), is taken to be a Gamma distribution, with specified parameters (see below). For computational efficiency, we use fecundity rather than viability selection; however, we ran a smaller number of simulations to verify that this choice does not lead to a detectable difference in the results. Each simulation is run for a burn-in period of $10N$ generations, to ensure convergence to the steady state behavior, before the variances at segregating sites are measured.

We explore a range of parameter values chosen to balance biological plausibility (see Section S4) and manageable running times. Notably, we used a population size of $N = 1000$, with a burn-in time of 10,000 generations. We vary the number of traits to include $n = 1, 3, \& 10$, and vary the mutation rate per haploid genome per generation within the range $\frac{1}{2N} \leq U \leq 1$ (see Section S4), including $U$=0.0005, 0.001. 0.002, 0.005, 0.01, 0.02, 0.05, 0.1, 0.2, 0.5 & 1. Selection coefficients are chosen from an exponential distribution (setting the shape parameter for the Gamma distribution to 1) with means E($S$) = 0.1, 10, and 50.

For simplicity, we take $w^2 = 1$, which is equivalent to choosing the units used to measure effect sizes.

The simulation results for the total genetic variance and its distribution among sites are in close agreement with our theoretical predictions (Fig. S6). Specifically, within the parameter ranges that we assume, i.e., when $1/2N \ll \sigma^2/w^2 \ll 1$ (see Section S4), the total genetic variances measured in simulations are indistinguishable from our predictions (Fig. S6A). Moreover, simulations and prediction seem to agree even when $\sigma^2/w^2 \lesssim 1/2N$, although we consider this range to be less relevant, given our focus on highly polygenic traits (see Section S4.4). We also compare simulated and predicted distributions of variances among sites, in terms of $G(v)$, the proportion of the variance arising from sites that contribute more than $v$ (Eq. S40), and found them to be in close agreement (Fig. S6B).

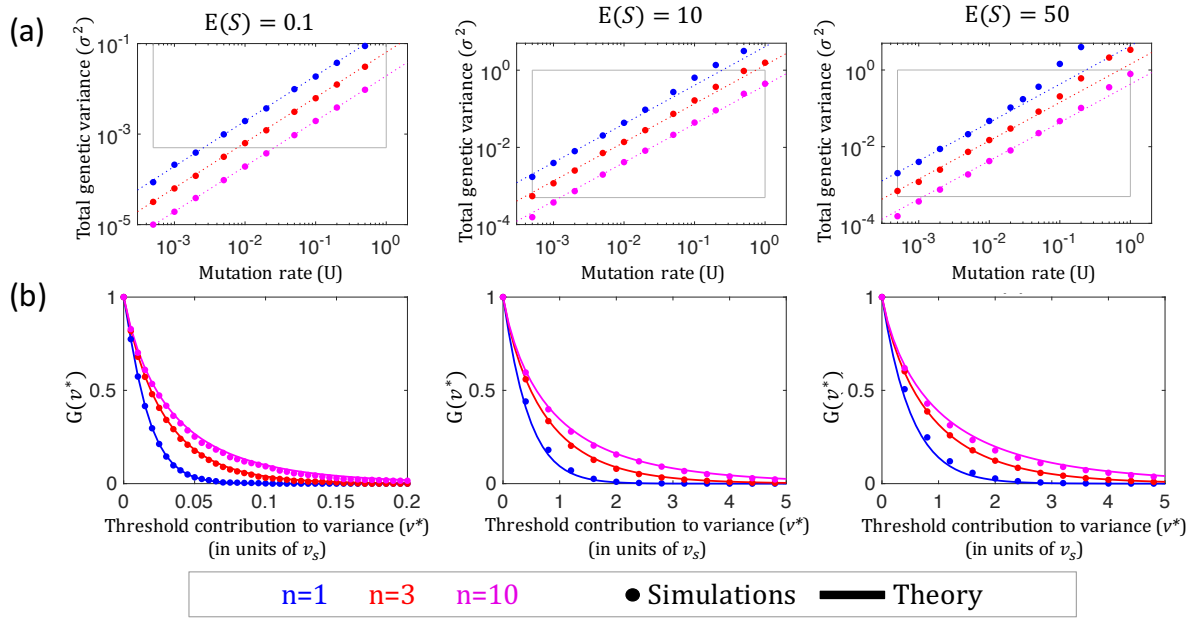

**Figure S6.** Testing theoretical predictions for genetic variance against simulation results. (a) Total genetic variance (in units of $w^2$) as a function of the mutation rate. The biologically relevant range of mutation rates, $1/2N \ll U \ll 1$, and the range in which we expect our predictions to be valid, $w^2/2N \ll \sigma^2 \ll w^2$, are marked by a grey box. (b) The distribution of variances among sites, for $U = 0.01$; $G(v^*)$ is the proportion of variance from sites contributing more than $v^*$ (Eq. S40). Error bars represent one standard deviation. For each set of parameters, the number of simulations was chosen to obtain standard deviations below 10%. In practice, we often obtain much smaller standard deviations $\ll 10\%$, which is why most error bars are too small to be visible.

We ran two additional variations on the basic simulation procedure (also available at https://github.com/sellalab/GenArchitecture): one to explore the effects of a shift in the optimal phenotype (Section S5.1 and Fig. S7) and another to explore the effects of asymmetric mutational input (Section S5.2 and Fig. S8). To these ends, for simplicity, we consider the case without pleiotropy, i.e., with $n = 1$. In the first, after the $10N$ generations burn-in period, we introduce a shift in the optimal phenotype, and trace the allelic behavior over an additional 4,000 generations (see Section S5.1). In the second, after the $10N$ generations burn-in period, we introduce asymmetry in the rates of trait increasing and decreasing mutations, and trace the allelic trajectories over an additional 10,000 generations. The parameters of these simulations are detailed in Sections S5.1 and S5.2, respectively.

## 4. Justification for assumptions

Here, we justify the assumptions that we relied upon in deriving the first two moments of change in allele frequency (see Section S2.1; modeling assumptions are motivated in the introduction to the main text). We rely in part on self-consistency arguments, which should not be mistaken for being circular: specifically, we make assumptions about the behavior of the system and show that the solution to which we arrive satisfies these assumptions.

### 4.1. Normal and isotropic phenotypic distribution around the optimum

The assumption that the phenotypic distribution is well approximated by a normal distribution stems from an additive model of quantitative traits. By assuming that the phenotype arises from many additive contributions and that these additive contributions arise from some underlying distribution, normality follows from the law of large numbers. In terms of model parameters, we would expect normality to hold if the rate of mutations affecting the trait is sufficiently large, i.e., when $2NU \gg 1$.

We further assume that the phenotypic distribution is isotropic and its mean is at the optimum. Isotropy of the phenotypic distribution follows from assuming isotropy in the mutational input. In Section S5.4, we explore the consequences of anisotropy in the mutational input. In Section S4.4, we further show that the fluctuations of the mean phenotype around the optimum over time have negligible effects on allelic dynamics; a similar argument applies to fluctuations in the variance.

### 4.2. The phenotypic variance satisfies $\sigma^2 \ll w^2$

With the mean phenotype centered at the optimum, requiring that $\sigma^2 \ll w^2$ is equivalent to assuming that moving a standard deviation away from the mean phenotype entails only a minor reduction in fitness. This seems plausible for many phenotypes: if, for example, this assumption did not hold for human height, then individuals whose height is a standard deviation or more away from the population mean would suffer from a substantial reduction in fitness. Arguably, deviations from the mean height would then be recognized as a very common and severe disease.

Another line of argument that it is likely that $\sigma^2 \ll w^2$ is based on our results. If we assume that mutations are strongly selected, then our results suggest that

$$\sigma^2 = 2NU \cdot v_s = \frac{4U}{n} w^2. \tag{S41}$$

It follows that if the rate of mutations affecting the phenotype under consideration satisfies $U \ll 1$ then $\sigma^2 \ll w^2$. The number of mutations per diploid human genome per generation is estimated to be ~60 (5), and less than 10% of the genome is assumed to be functional (6), suggesting that the number of de novo mutations with any effect on function is less than 3 per haploid per generation. It then seems plausible that the (haploid) mutation rate affecting a specific trait satisfies $U \ll 1$. Assuming that mutations are weakly selected increases the variance in Eq. S41 only moderately and assuming the mutations are effectively neutral would suggest it is much smaller, leaving the above argument intact.

### 4.3. Mutational effect sizes satisfy $a^2 \ll w^2$

As we argued in the introduction of the main text, variants for which the stronger condition $a^2 \ll \sigma^2$ holds account for most or all of the heritability explained in GWAS for many traits (e.g., 7, 8-10). Moreover, evidence for many traits suggests that the same is true for the variants that underlie the heritability that remains to be explained (11-14). Indeed, for this assumption to be violated, much of the genetic variance would have to arise from mutations that have a very large impact on fitness (i.e., with *s* on the order of 1). While this may be the case for some diseases (e.g., autism (15)), it does not appear to be the case for most phenotypes that have been examined.

### 4.4. Deviations of the mean phenotype from the optimum can be neglected

In reality, the mean phenotype of the population fluctuates around the optimum. Here, we derive equations for the dynamic of the mean phenotype in order to estimate the magnitude and timescale of these fluctuations. We then show that these fluctuations have a negligible effect on the first two moments of change in allele frequency and thus on the results that follow from these moments.

We begin by deriving the first and second moment of change in mean phenotype. To this end, we assume the distribution of phenotypes is a multivariate normal centered around a mean phenotype, $\bar{r}$, i.e. that

$$f(\vec{r}) = \frac{1}{(2\pi\sigma^2)^{n/2}} \exp\left(-\frac{|\vec{r}-\bar{r}|^2}{2\sigma^2}\right). \tag{S42}$$

The expected change in mean phenotype due to selection in one generation is therefore

$$E(\Delta \bar{r}) = \frac{\int_{\vec{r}} f(\vec{r}) W(\vec{r}) \vec{r}}{\int_{\vec{r}} f(\vec{r}) W(\vec{r})} - \bar{r} = -\frac{\sigma^2}{w^2+\sigma^2}\bar{r} \approx -\frac{\sigma^2}{w^2}\bar{r}. \tag{S43}$$

By the same token, the variance in $\Delta \bar{r}$ is simply the sampling variance

$$V(\Delta \bar{r}) \approx \frac{\sigma^2}{N}, \tag{S44}$$

where in both cases we relied on the assumption that $\sigma^2 \ll w^2$.

These two moments define an Ornstein-Uhlenbeck process in $\bar{r}$, allowing us to rely on well-known results (16). Notably, when the mean phenotype $\bar{r}$ starts far from the optimum, it decays exponentially to the optimum with exponent $\sigma^2/w^2$ (see Section S5.1 below). At steady state, $\bar{r}$ will fluctuate with mean zero and $E(\bar{r}^2) = nw^2/2N$ over a time scale of $w^2/\sigma^2$ generations. The typical displacement of $\bar{r}$ in any given direction will be $\sqrt{w^2/2N}$, reflecting a balance between drift and the pull of selection toward the optimum.

Next we show that these fluctuations of the mean have negligible effects on allelic trajectories. To this end, we derive the first two moments of change in allele frequency, but this time, we include the effect of the displacement of $\bar{r}$ from the optimum. While the second moment remains the same, the first moment becomes

$$E(\Delta q) \approx -\frac{\bar{r} \cdot \vec{a}}{w^2} pq - \frac{a^2}{w^2} pq \left(\frac{1}{2} - q\right) = -\frac{\bar{r}_a}{\sqrt{w^2/2N}} \frac{\sqrt{S}}{2N} pq - \frac{S}{2N} pq \left(\frac{1}{2} - q\right), \tag{S45}$$

where $\bar{r}_a$ is $\bar{r}$'s component in the direction of $\vec{a}$. However, our analysis establishes that $\frac{\bar{r}_a}{\sqrt{w^2/2N}}$ is a scalar of the order of 1, which fluctuates around zero on a timescale of $w^2/\sigma^2$. We can therefore compare the first term in the above equation, which represents directional selection, and the second term, which represents stabilizing selection. When stabilizing selection is strong, $S \gg 1$, the stabilizing selection term dominates over the directional selection term. In contrast, when selection is weak, i.e., $S \approx 1$ or smaller, then in any given generation, the directional term is not necessarily negligible. However, in this case, both terms affect substantial change in allele frequency only over a timescale of $2N$ generations; on this timescale, if $2N \gg w^2/\sigma^2$, the directional effect would average to zero. The directional term will become important only when $2N \leq w^2/\sigma^2$, that is $\sigma^2 \leq w^2/2N$. For $\sigma^2$ to be that small, virtually all alleles must have $S \ll 1$, such that their trajectories will be determined by drift, not selection. In summary, regardless of the selection acting on an allele, fluctuations of the mean phenotype around the optimum will have a negligible effect on its trajectory.

## 5. Model robustness

In this section, we consider the sensitivity of our results to relaxing some of the simplifying modeling assumptions about selection and mutation. Specifically, we show our results to be robust to moderate changes to the optimal phenotype; small asymmetry in the mutational input; the presence of major loci maintained at high frequency by selection on traits that are not included in the model; as well as to most forms of anisotropic mutation.

### 5.1. Changes to the optimal phenotype

We first consider how changes to the optimal phenotype over time would affect our results. It is easy to imagine how events such as migration from Africa to Europe or the onset of agriculture may have introduced rapid changes in optimal phenotypes. In order to evaluate the potential impact of such events, we consider how an instantaneous change to the optimal phenotype would affect the allelic dynamics. Similar models have recently been analyzed in the limit of infinite population size (17, 18).

We begin by considering how such an instantaneous change to the optimum would affect the mean phenotype. If the shift to the optimum is small, on the order of the fluctuations in the mean phenotype at steady state or smaller, then the arguments provided in Section S4.4 will still hold and the shift would have a negligibly small effect on our results. We therefore assume that the shift in optimum, $\vec{z}$, is large compared to the scale of fluctuations ($z^2 \gg w^2/2N$). This assumption means that we can use a deterministic approximation (based on Eq. S43) and describe the change in mean phenotype in a single generation by

$$\Delta \vec{r} \approx \mathrm{E}(\Delta \vec{r}) = -\frac{\sigma^2}{w^2}(\vec{r} - \vec{z}) \qquad (S46)$$

(neglecting higher moments). Further assuming that the mean phenotype was at the optimum, $\vec{0}$, before the optimum shifted (at time $t = 0$) and neglecting changes to the genetic variance $\sigma$, we find that

$$\vec{r}(t) = \vec{z}\left(1 - \exp\left(-\frac{\sigma^2}{w^2}t\right)\right) \qquad (S47).$$

Thus, the mean $\vec{r}$ adapts to the new optimum on a timescale of $w^2/\sigma^2$ generations (see (19) for a similar derivation).

We can rely on this approximation to learn when a shift in optimum will have negligible effects on allele trajectories. Recalling Eq. S45, the first moment of change in allele frequency is given by

$$\mathrm{E}(\Delta q) \approx -\frac{(\vec{r}(t)-\vec{z})\cdot\vec{a}}{w^2}pq - \frac{a^2}{w^2}pq\left(\frac{1}{2}-q\right), \tag{S48}$$

where, based on our approximation (Eq. S47), the directional selection term introduced by the shift in optimum takes the time-dependent form

$$\mathrm{E}(\Delta_D q) = -\frac{(\vec{r}(t)-\vec{z})\cdot\vec{a}}{w^2}pq \approx \frac{\vec{z}\cdot\vec{a}}{w^2}\exp\left(-\frac{\sigma^2}{w^2}t\right)pq. \tag{S49}$$

The effect of this directional term over the entire adaptive trajectory can be quantified by comparing the expected allele frequency after adaptation to the shift, $q_D$, with initial frequency before the shift, $q_0$, i.e.,

$$\ln(q_D/q_0) = \int_t \frac{\mathrm{E}(\Delta_D q)}{q(t)} = \frac{\vec{z}\cdot\vec{a}}{w^2}\int_t p(t)\exp\left(-\frac{\sigma^2}{w^2}t\right) < \frac{\vec{z}\cdot\vec{a}}{w^2}\int_t \exp\left(-\frac{\sigma^2}{w^2}t\right) = \frac{\vec{z}\cdot\vec{a}}{\sigma^2}. \tag{S50}$$

This result suggests that the relative change in allele frequency will be negligible so long as

$$(\vec{z}\cdot\vec{a})/\sigma^2 \ll 1. \tag{S51}$$

This condition suggests that mutations with smaller effects would be less affected by the shift in the optimum. It further suggests that alleles that satisfy $a^2 \ll \sigma^2$, as appears to be the case for most loci discovered in GWAS (e.g., 7, 8-10), will be negligibly affected by shifts in optimum on the order of the total genetic variation (i.e., $z \leq \sigma$). These analytic predictions are confirmed by simulations (Fig. S7).

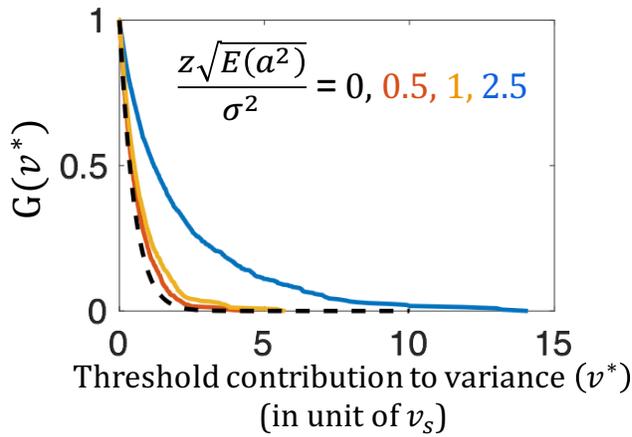

**Figure S7.** Distribution of the contributions of sites to variance after a shift in the optimum. The y-axis is the proportion of the variance explained by sites that contribute more than $v^*$ to the variance. The theoretical prediction without adaptation is shown in dashed black, and simulation results for different shifts in the optimal phenotype are shown in color. When the root mean square of $(\vec{z}\cdot\vec{a})/\sigma^2$ becomes larger than 1, directional selection substantially affects allele frequencies and therefore the contributions of sites to variance, as predicted by Eq. S51. (Since mutation is symmetric the mean of $(\vec{z}\cdot\vec{a})/\sigma^2$ is zero and we quantify its characteristic value by its root mean square $z\sqrt{E(a^2)}/\sigma^2$.) Simulations were run with an exponential distribution of selection coefficients with $E(S) = 25$, $N = 1,000$, $n = 1$, $U = 0.01$, and a burn-in time of 10,000 generations. Results were taken 50 generations after the shift in optimum, which, for these parameters, is just after the population mean has reached the new optimum.

## 5.2. Asymmetric mutational input

In this section, we consider he sensitivity of our results to asymmetries in the mutational input, i.e., to the case in which mutations in a given direction in trait space, $\vec{b}$, are more likely to arise than mutations in the opposite direction, $-\vec{b}$ (see (20) for treatment of this problem in the limit of high per-site mutation rate).

An asymmetric mutational input introduces a shift in the mean phenotype every generation. With new mutations arising at frequency $1/2N$, the expected shift is

$$\Delta_M \bar{r} = 4NU\, E_M(\vec{a}) \cdot 1/2N = 2U E_M(\vec{a}), \tag{S52}$$

where $E_M$ is the expectation over newly arising mutations. For each trait, effects have a characteristic size $\sqrt{E(a^2)/n} = \sqrt{v_s}\sqrt{E(S/4)}$. The characteristic effect size sets the scale for the maximal shift in any direction, that is $\|\Delta_M \bar{r}\|$ is of the order of $2U\sqrt{E(a^2)/n}$ or smaller. We therefore parameterize the shift in mean phenotype due to new mutations by

$$\Delta_M \bar{r} = 2U E_M(\vec{a}) = 2U\sqrt{E(a^2)/n}\,\vec{b}, \tag{S53}$$

where the vector $\vec{b}$ parameterizes the strength and direction of the bias and $b = |\vec{b}|$ is assumed to be $\ll 1$.

At steady state, the mutational shift must be offset by selection, such that

$$\Delta_M \bar{r} + \Delta_D \bar{r} + \Delta_S \bar{r} = \vec{0}, \tag{S54}$$

where $\Delta_D \bar{r}$ and $\Delta_S \bar{r}$ are the expected shifts due to directional and stabilizing selection, respectively. We previously found that the expected directional shift is

$$\Delta_D \bar{r} = -\frac{\sigma^2}{w^2}\bar{r}, \tag{S55}$$

where $\bar{r}$ denotes the mean phenotype (see Eq. S43). As we show next, when mutations are strongly selected, stabilizing selection offsets the mutational shift to maintain the mean phenotype at the optimum, implying that directional selection is negligible. In contrast, when mutations are effectively neutral, stabilizing selection is negligible and a directional term might not be negligible by comparison. However, as long as asymmetry is small, $b \ll 1$, we show that this directional term is not large enough to change the allele dynamics, both when all mutations are effectively neutral and when some mutations are strongly selected.

First, we consider the shift in mean phenotype due to stabilizing selection. This shift arises because, with asymmetric mutational input, the distribution of phenotypes becomes skewed. Therefore, even if the mean phenotype is at the optimum, individuals with a given fitness may have an asymmetric distribution of phenotypes around the optimum, leading stabilizing selection to change the mean phenotype. We have already shown (Eq. S15) that the expected change in allele frequencies per generation due to stabilizing selection at any given site $i$ is

$$E(\Delta q_i) = -\frac{a_i^2}{w^2} p_i q_i \left(\frac{1}{2} - q_i\right). \tag{S56}$$

The expected change in mean phenotype can then be calculated by adding up the contributions over sites

$$\Delta_S \bar{r} = -E\left(\sum_i 2\vec{a}_i \frac{a_i^2}{w^2} p_i q_i \left(\frac{1}{2} - q_i\right)\right). \tag{S57}$$

The right-hand side of this equation reflects the skewness of the phenotypic distribution. Indeed, in one dimension, it can be shown that

$$\Delta_S \bar{r} = -\frac{\mu_3(r)}{2w^2}, \tag{S58}$$

with $\mu_3(r)$ being the third central moment of the phenotypic distribution. In $n$-dimensions, for every direction $x$,

$$\Delta_S \bar{r}_x = -\frac{1}{2w^2} E((\vec{r} - \bar{r})_x (\vec{r} - \bar{r})^2) = -\frac{1}{2w^2} \sum_k \mu_3(\vec{r})_{xkk}, \tag{S59}$$

with $\mu_3(\vec{r})_{jkl} = E\big((\vec{r} - \bar{r})_j (\vec{r} - \bar{r})_k (\vec{r} - \bar{r})_l\big)$.

When sites are under strong selection, $\Delta_S \bar{r}$ takes a simple form. Assuming the asymmetry is small, the shift due to stabilizing selection can be expanded in orders of $b$. The leading term in the frequency distribution takes the same form as it does without the bias. For strongly selected alleles with no bias, $q \ll 1$ and therefore the frequency dependence in this term can be approximated by $pq \left(\frac{1}{2} - q\right) \approx \frac{1}{2} q$. Moreover, $q$ scales with $1/a^2$, implying that the distribution of $a^2 q$ is independent of $\vec{a}$ and that $E(a^2 q) = w^2/N$ (see Section S3.1). Therefore, when all sites are strongly selected, the leading term in the shift due to stabilizing selection is

$$\Delta_S^0 \bar{r} = -E\left(\sum_i 2\vec{a}_i \frac{a_i^2}{w^2} \frac{q_i}{2}\right) = -\frac{E(a^2 q)}{w^2} E(\sum_i \vec{a}_i) = -\frac{1}{N} 2NU E_M(\vec{a}) = -U\sqrt{v_s}\sqrt{E(S)}\, \vec{b}$$

$$= -\Delta_M \bar{r}. \tag{S60}$$

Thus, to a first order in $b$, the shift of the mean phenotype due to stabilizing selection offsets the mutational shift, implying that there will be no directional term and that the allele dynamics will not be affected by asymmetry.

When alleles are instead effectively neutral, then $a^2/w^2 \ll 1/2N$ (see Section S2.2) and allele frequencies are well approximated by the neutral sojourn time, $\tau(q) \approx 2/q$. The shift due to stabilizing selection then satisfies

$$\Delta_S \bar{r} = -\mathrm{E}\left(\sum_i 2\vec{a}_i \frac{a_i^2}{w^2} p_i q_i \left(\frac{1}{2} - q_i\right)\right) \approx -\mathrm{E}\left(pq\left(\frac{1}{2} - q\right)\right) \mathrm{E}\left(\sum_i 2\vec{a}_i \frac{a_i^2}{w^2}\right)$$

$$= -\frac{1}{6}\mathrm{E}\left(\sum_i 2\vec{a}_i \frac{a_i^2}{w^2}\right) \ll -\frac{1}{N}\mathrm{E}(\sum_i \vec{a}_i) = -\Delta_M \bar{r}, \tag{S61}$$

implying that it makes a negligible contribution to offsetting the mutational shift. In this case, the mutational effect on the mean phenotype is therefore offset by directional selection, where

$$\Delta_D \bar{r} = -\frac{\sigma^2}{w^2}\bar{r} \approx -\Delta_M \bar{r}, \tag{S62}$$

indicating a displacement of the mean phenotype from the optimum

$$\bar{r} = \frac{w^2}{\sigma^2}\Delta_M \bar{r}. \tag{S63}$$

This displacement introduces a directional selection term into the first moment of change in allele frequency that, if large enough, could alter allele dynamics (see Section S4.4). However, when all alleles are effectively neutral, we have

$$\bar{r} = \frac{w^2}{\sigma^2}\Delta_M \bar{r} = \frac{w^2}{2NUv_s\mathrm{E}(S)/2} 2U\sqrt{v_s}\sqrt{\mathrm{E}(S/4)}\,\vec{b} = \frac{1}{2N}\frac{2w^2}{\sqrt{v_s}\sqrt{\mathrm{E}(S)}}\,\vec{b}, \tag{S64}$$

and therefore the scaled directional selection coefficient, for an allele with effect size $\vec{a}$ and scaled stabilizing selection coefficient $S = 2N\frac{a^2}{w^2}$, is of the order of

$$2N\frac{\vec{r}\cdot\vec{a}}{w^2} = \frac{2b_a a}{\sqrt{v_s}\sqrt{\mathrm{E}(S)}} \sim \frac{2(b/\sqrt{n})a}{\sqrt{v_s}\sqrt{\mathrm{E}(S)}} = \frac{b\sqrt{v_s}\sqrt{S}}{\sqrt{v_s}\sqrt{\mathrm{E}(S)}} = \sqrt{\frac{S}{\mathrm{E}(S)}}\,b, \tag{S65}$$

with $b_a \sim b/\sqrt{n}$ being the projection of $\vec{b}$ in the direction of $\vec{a}$. Since $b \ll 1$, for all alleles other than those with unusually large selection coefficients, the scaled directional selection coefficient will be much smaller than 1 and the trajectories will still be determined by drift and not selection. Even in this case, therefore, we do not expect asymmetry to affect allele dynamics.

Next, we consider the case where there is a mix of effectively neutral and strongly selected mutations. The existence of strongly selected mutations in addition to effectively neutral ones reduces the deviation of the mean phenotype from the optimum. Denoting the proportion of strongly selected mutations by $p_s$, we have

$$\bar{r} = \frac{w^2}{\sigma^2} U(1-p_s)\sqrt{v_s}\sqrt{\mathrm{E}(S^{e.n.})}\,\vec{b}, \tag{S66}$$

where $\mathrm{E}(S^{e.n.}) \leq 1$ is the mean scaled stabilizing selection coefficient for effectively neutral mutations. Since $\sigma^2 > 2NUp_s v_s$, we can then obtain an upper bound to the magnitude of scaled directional selection coefficient for an allele with effect size $\vec{a}$ and scaled stabilizing selection coefficient $S = 2N\frac{a^2}{w^2}$:

$$2N\frac{\bar{r}\cdot\vec{a}}{w^2} = 2N\frac{1}{\sigma^2} U(1-p_s)\sqrt{v_s}\sqrt{\mathrm{E}(S^{e.n.})}\,\vec{b}\cdot\vec{a}$$
$$< \frac{1}{Up_s v_s} U(1-p_s)\sqrt{v_s}\sqrt{\mathrm{E}(S^{e.n.})}\,\vec{b}\cdot\vec{a} \sim \frac{1-p_s}{2p_s}\sqrt{\mathrm{E}(S^{e.n.})}\sqrt{S}\,b. \tag{S67}$$

With a substantial proportion of strongly selected sites, $(1-p_s)/2p_s$ is of the order of 1, and therefore $\frac{1-p_s}{2p_s}\sqrt{\mathrm{E}(S^{e.n.})}\,b \ll 1$. This condition implies that for effectively neutral alleles (i.e., $S \leq 1$), the scaled directional selection coefficient is $\ll 1$ and allele trajectories will be determined by genetic drift, whereas for strongly selected alleles (i.e., when $S \gg 1$), the scaled directional selection coefficient is $\ll S$ and therefore negligible compared to the scaled stabilizing selection coefficient.

Weakly selected alleles (with $1 < S < 30$) behave largely like strongly selected alleles except that stabilizing selection on them only partially cancels out the mutational bias (for example, for $S = 10$ only 85% of the bias is canceled). The rest of the bias is canceled by directional selection and therefore induces a small shift in the mean phenotype. It is straightforward to repeat the arguments given above and show that the shift in the mean phenotype for a trait with only weakly selected alleles or a mixture that includes weakly selected alleles negligibly affects allele trajectories.

Thus, we conclude that small asymmetry in mutation will not affect the allelic dynamic (see Fig. S8).

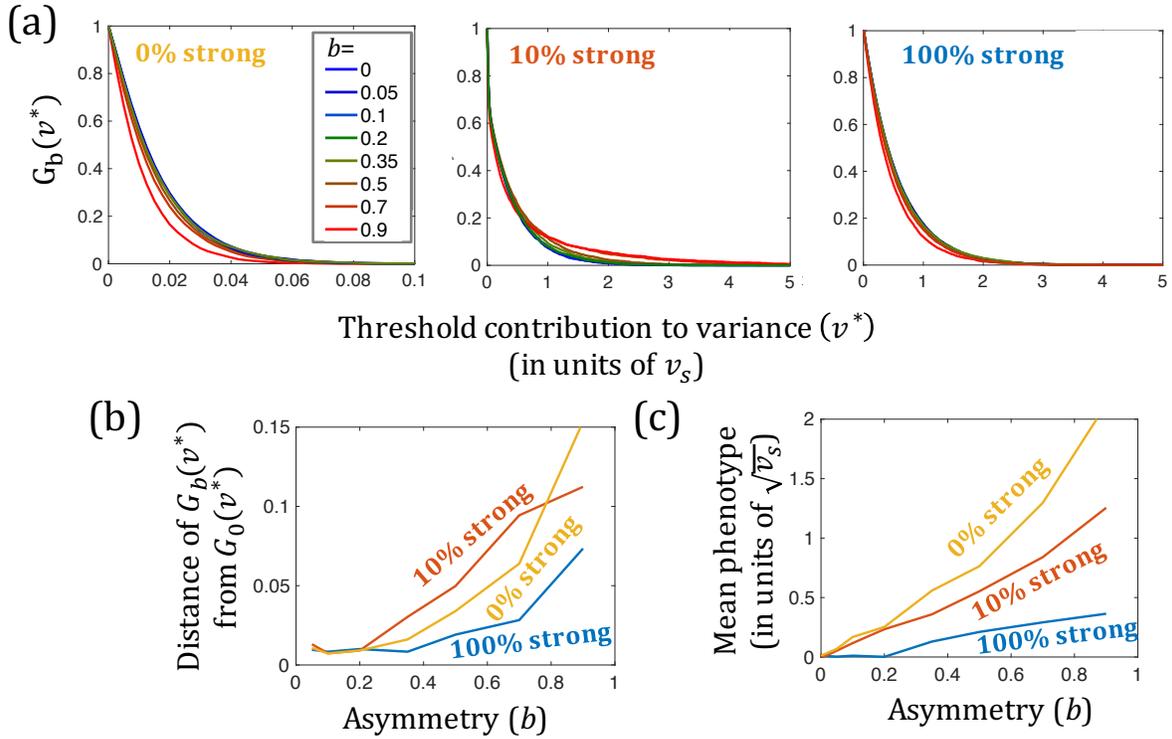

**Figure S8.** The effect of asymmetric mutational input on the contribution of sites to variance and the mean phenotype. (a) Proportion of genetic variance as a function of the threshold contribution to variance $v^*$, i.e., $G_b(v^*)$, for different bias strengths. (b) The maximal distance of $G_b(v^*)$ from $G_0(v^*)$, i.e. $\arg\max_{v^*}(|G_b(v^*) - G_0(v^*)|)$, as a function of $b$. (c) The mean phenotype $\bar{r}$, in units of $\sqrt{v_s}$, as a function of mutational bias $b$. Simulations were run with $N = 1,000$, $n = 1$ and with different mixtures of effectively neutral (distributed around $S = 0.1$) and strong (distributed around $S = 50$) selection coefficients. Asymmetry was simulated by having more trait increasing than trait decreasing mutations; if $\beta$ is the proportion of trait increasing mutations then the asymmetry coefficient is $b = 2\beta - 1$. As expected, for small biases (when $b \ll 1$), there are no substantial changes in the distribution of the contribution of sites to variance. Simulations were run with a 10,000 generations burn-in period without asymmetry and then 10,000 generations with asymmetry and averaged over many runs (>300), with the number of runs varied across plots keep errors in (a) below 1%.

## 5.3. Major effect loci

In this section, we show that our results are insensitive to the presence of major loci, i.e., individual loci that contribute substantially to quantitative genetic variation. We have in mind, for example, loci whose alleles are maintained at high frequencies by balancing selection on a Mendelian trait but have pleiotropic effects on the quantitative traits under consideration (e.g., HLA loci; (21, 22)). While such loci violate our assumptions, we show that they do not affect the dynamics at other loci that fulfill them.

To this end, we calculate the first two moments of change in allele frequency in the presence of a major locus. We denote the frequency and effect size of the focal allele by $q$ and $\vec{a}$, and the frequency and effect size of the major allele by $q_M$ and $\vec{a}_M$, respectively. As in our previous derivations (Section S2.1), the distribution of background phenotypic contribution from all other loci, $\vec{R}$, is well approximated by the normal distribution

$$f(\vec{R}|\vec{a}_M, q_M, \vec{a}, q) = \frac{1}{(2\pi(\sigma^2 - \sigma_M^2))^{n/2}} \exp\left(-\frac{(\vec{R} + 2q\vec{a} + 2q_M\vec{a}_M)^2}{2(\sigma^2 - \sigma_M^2)}\right), \quad (S68)$$

where $\sigma_M^2$ is the contribution to genetic variance from the major locus. The population mean remains close to the optimum because any shift caused by the major locus is quickly compensated for by the other loci (see Section S4.4). We then average over both this distribution and the three genotypes at the major locus to calculate the mean fitness associated with each genotype at the focal locus. Namely,

$$W_{00} = (1 - q_M)^2 \int_{\vec{R}} f(\vec{R}|\vec{a}_M, q_M, \vec{a}, q) W(\vec{R}) + 2q_M(1 - q_M) \int_{\vec{R}} f(\vec{R}|\vec{a}_M, q_M, \vec{a}, q) W(\vec{R} + \vec{a}_M)$$

$$+ q_M^2 \int_{\vec{R}} f(\vec{R}|\vec{a}_M, q_M, \vec{a}, q) W(\vec{R} + 2\vec{a}_M), \quad (S69)$$

and similarly for the other genotypes. In this way, we obtain the first moment of the change in allele frequency

$$E(\Delta q) = -pq \frac{p(W_{00} - W_{01}) + q(W_{01} - W_{11})}{\overline{W}} \approx -\frac{a^2}{w^2} pq\left(q - \frac{1}{2}\right), \quad (S70)$$

which is the same as we derived in the absence of a major locus (Eq. S15). Similarly, we find the second moment to be unaffected.

### 5.4. Anisotropic mutation

In this section, we consider how relaxing the assumption that the distribution of newly arising mutations is isotropic in trait space would affect our results. As noted, we can always choose an orthonormal coordinate system centered at the optimum, in which the trait under consideration varies along the first coordinate and a unit change in other traits (i.e., in other coordinates) near the optimum have the same effect on fitness. There is, however, no obvious reason for the distribution of newly arising mutations to be isotropic in this coordinate system (see (23) for generalizations of Fisher's Geometric Model along similar lines).

Anisotropy in mutation does not affect the moments of change in allele frequency, as these depend only on the selection on an allele or equivalently on its effect size but not on its direction in trait space. Anisotropy could affect the distribution of allelic effect sizes on the focal trait conditional on the selection acting on them. Here, we provide heuristic arguments suggesting that, barring extreme cases, we can define an effective number of traits $n_e$ and an effective strength of selection $w_e^2$ for which the relationship between selection and effect size in anisotropic models is well approximated by the relationship found for isotropic ones (Eqs. 9 & 11; Section S1.2).

We focus on a family of anisotropic mutational distributions that can be described as a projection of a multivariate normal distribution on the unit sphere in trait space. Namely, we draw the size of a mutation $a = \|\vec{a}\|$ from some distribution and to obtain its direction, we draw a vector $\vec{\alpha}$ from a multi-variate normal distribution $\text{MVN}(0, \boldsymbol{\Sigma})$ and normalize it, i.e.,

$$\vec{a} = a \frac{\vec{\alpha}}{\alpha}, \tag{S71}$$

and therefore

$$a_1 = a \frac{\alpha_1}{\alpha}. \tag{S72}$$

This family of mutational distributions gives us a mathematically tractable framework with which to examine the behavior of our model with anisotropy.

With anisotropy, the behavior of our model greatly depends on the relative contribution of the focal trait to selection, which we parameterize by

$$\gamma_1 \equiv \frac{\text{E}(\alpha_1^2)}{\text{E}(\alpha^2)} = \frac{\Sigma_{11}}{\text{tr}(\boldsymbol{\Sigma})}. \tag{S73}$$

When selection acts mainly on our focal trait, i.e. when $\gamma_1 \approx 1$, then $|\alpha_1| \approx \alpha$ and therefore $a_1 \approx \pm a$. Such a relationship between the strength of selection and effect size is well approximated by an isotropic model with $n_e = 1$. We therefore focus on cases in which there is a significant pleiotropic contribution to selection, i.e., $\gamma_1$ is substantially less than 1. Anisotropy then has two effects: the first is to introduce heterogeneity in the strength of selection on different traits and the second is to introduce correlations in the effects of a mutation on different traits, notably between the focal trait and others.

We first consider the case in which the strength of selection differs among traits, but traits are uncorrelated, corresponding to a diagonal covariance matrix, $\boldsymbol{\Sigma}$. When many traits have a non-negligible contribution to selection, $\alpha^2 = \alpha_1^2 + \alpha_2^2 + \cdots \alpha_n^2$ would have a small coefficient of variation, i.e., $C_V^2(\alpha^2) = V(\alpha^2)/E(\alpha^2)^2 \ll 1$, because of the law of large numbers. In this case,

$$a_1 = a\frac{\alpha_1}{|\vec{\alpha}|} = a\frac{\alpha_1}{\sqrt{E(\alpha^2)}}\left(1 + O\left(C_V^2(\alpha^2)\right)\right) \approx a\frac{\alpha_1}{\sqrt{E(\alpha^2)}} = \frac{a}{\sqrt{1/\gamma_1}} \frac{\alpha_1}{\sqrt{E(\alpha_1^2)}}, \tag{S74}$$

Since $\alpha_1/\sqrt{E(\alpha_1^2)} \sim N(0,1)$ and $s = \frac{1}{w^2}a^2$ this implies that, conditional on the selection coefficient $S$, the effect size on the focal trait will be distributed as

$$a_1 \sim N\left(0, \frac{w^2}{n_e}s\right) \tag{S75}$$

with $n_e = 1/\gamma_1$. This is the same relationship between selection and effect size as the high pleiotropy isotropic model with $n = n_e$ (Eq. 11 of main text). This result suggests the concept of an effective number of traits, which can be thought of as the number of traits that have the same effect on fitness as the focal one and are required to produce the same strength of selection on alleles. The effective number of traits describes the distribution of effect sizes both in the limit of high pleiotropy $n_e \gg 1$ and low pleiotropy $n_e \approx 1$ and simulations show that it describes the distribution, at least qualitatively, also for intermediate values of $n_e$ (Fig. S9).

However, there is an extreme scenario in which an effective number of traits cannot describe the distribution of effect sizes. This happens when $C_V^2(\alpha^2) \geq 1$, that is when selection acts mainly on a small number of traits but our focal trait contributes very little to selection ($\gamma_1 \ll 1$). In this case, we might be tempted to use $n_e = 1/\gamma_1 \gg 1$ but, as Eq. S74 suggests, the high pleiotropy limit would be inadequate. In fact, the variance in selection on newly-arising mutations (due to the contribution of the selected traits) will result in a long-tailed distribution of effect sizes on the focal trait, which is not well-approximated by any isotropic model. In summary, except for these extreme cases, isotropic models provide a good approximation for the relationship between selection and effect size, even when there is heterogeneity in the strength of selection on different traits.

To illustrate the effect of heterogeneity in the strength of selection among traits, we consider a simple example in which all non-focal traits make the same contribution to selection and therefore can be modeled by

$$\Sigma = \begin{pmatrix} 1 & 0 & 0 & \cdots \\ 0 & c^2 & 0 & \\ 0 & 0 & c^2 & \\ \vdots & & & \ddots \end{pmatrix}, \tag{S76}$$

with $c^2$ being the ratio between the expected contribution of a non-focal trait to selection and the expected contribution of the focal trait to selection. It is easy to see that in this model $n_e = 1/\gamma_1 = 1 + (n-1)c^2$. Numerical results for this model are shown in Fig. S9.

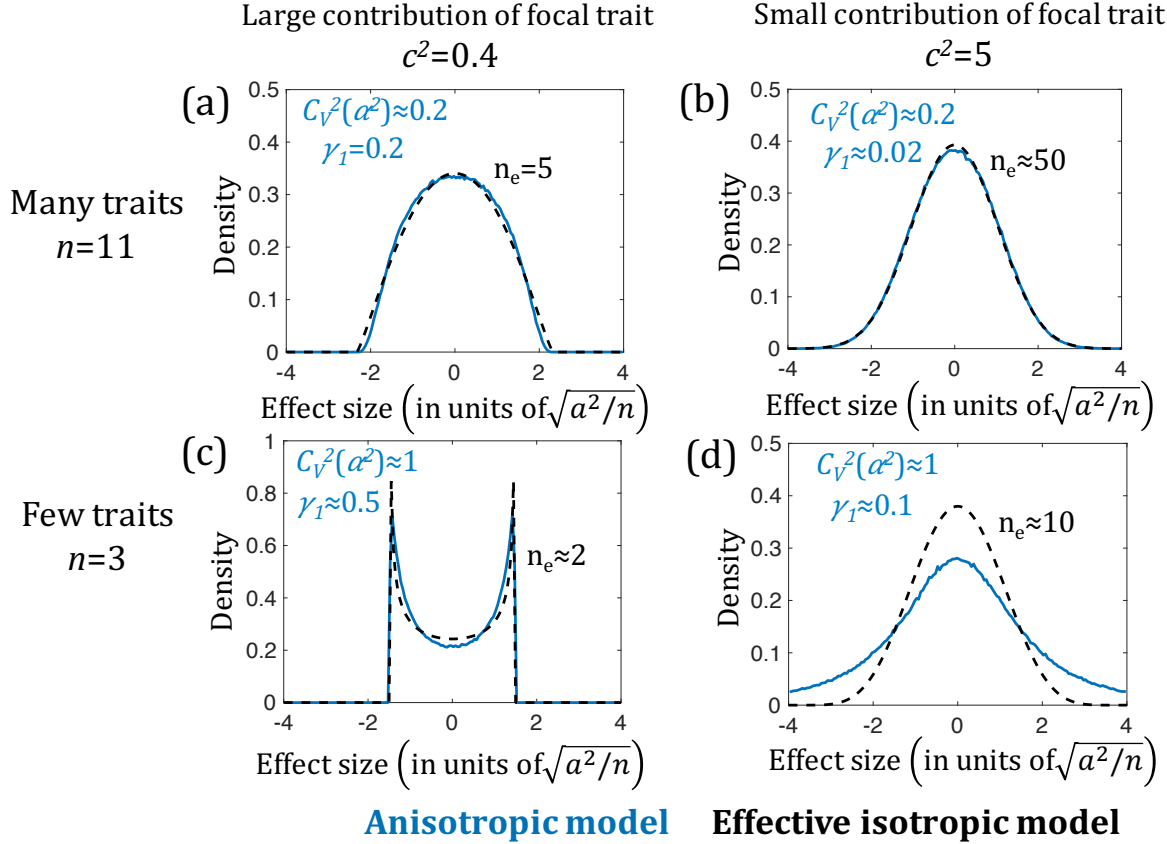

**Figure S9.** The effects of heterogeneity in the strength of selection on different traits on the distribution of effect sizes in the focal trait. Numerical results for models with the correlation matrix defined in Eq. S76 are shown in blue and the corresponding isotropic model in black dashes. When there are many selected traits, an isotropic model with $n_e = 1/\gamma_1 = 1 + (n-1)c^2$ provides a good approximation of the distribution of effect sizes, both when the focal trait contributes substantially to selection (a) and when it does not (b). When there are few traits, an isotropic model with $n_e = 1/\gamma_1 = 1 + (n-1)c^2$ provides a good approximation only when the focal trait contributes substantially to selection (c & d).

Next, we consider the case in which the effect sizes on different traits are correlated, i.e., when the covariance matrix $\Sigma$ has off-diagonal terms. $a_1^2 \propto \alpha_1^2/\alpha^2$ and therefore we parameterize the effect of these terms using the correlation between $\alpha_1^2$ and $\alpha^2$, $\rho^2 \equiv \text{corr}(\alpha^2, \alpha_1^2)$. If the correlation is small, $\rho^2 \ll 1$, then our previous reasoning holds. In the other extreme, when all selected traits are highly correlated with the focal trait, i.e. $\rho^2 \approx 1$, then the proportional contribution of the focal trait to selection is constant, $\alpha_1^2/\alpha^2 \approx \mathrm{E}(\alpha_1^2)/\mathrm{E}(\alpha^2) = \gamma_1$, and the effect size is $a_1 = \pm\sqrt{\gamma_1}\, a$. This model is therefore equivalent to an isotropic one with $n_e = 1$ and $w_e^2 = \gamma_1 w^2$; the latter change corresponds to increasing the strength of selection on the focal trait to account for selection on the other traits which are highly correlated with the focal trait. Intermediate cases are more complex: while effect sizes are still of the order of $\sqrt{\gamma_1}\, a$, the shape of the distribution of effect sizes is intermediate between the single trait and high pleiotropy limits. Isotropic models with an effective number of traits, $n_e < 1/\gamma_1$, and increased selection $w_e^2 = \gamma_1 n_e w^2$ can qualitatively describe these cases but may not completely capture the distribution of effect sizes. The value of $n_e$ would change from 1 when $\rho^2 \to 1$ to $1/\gamma_1$ when $\rho^2 \to 0$. Note that with a large number of traits, very strong correlations among many of the traits will be necessary in order to create a large enough $\rho^2$ to have a significant effect on $n_e$ (see Fig. S10).

To illustrate the effect of correlations among traits, we consider the following simple example (Fig. S10). We assume the correlation matrix $\Sigma$ takes the form

$$\Sigma = \begin{pmatrix} 1 & r^2 & r^2 & \cdots \\ r^2 & 1 & r^2 & \\ r^2 & r^2 & 1 & \\ \vdots & & & \ddots \end{pmatrix}, \tag{S77}$$

meaning that that all traits contribute equally to the fitness and every pair of traits has the same correlation coefficient $r^2$. When $r^2 = 0$ this becomes an isotropic model. When $r^2 = 1$, effect sizes are always identical for every trait; thus, this case is equivalent to having only one trait with selection that is increased $n$-fold. Intermediate cases can be approximated by finding an effective number of traits $n_e < 1/\gamma_1 = n$, such that an isotropic model with $n_e$ and $w_e^2 = \gamma_1 n_e w^2 = w^2 n_e/n$ qualitatively describes the distribution of effect sizes. Numerical results of this model are shown in Fig. S10.

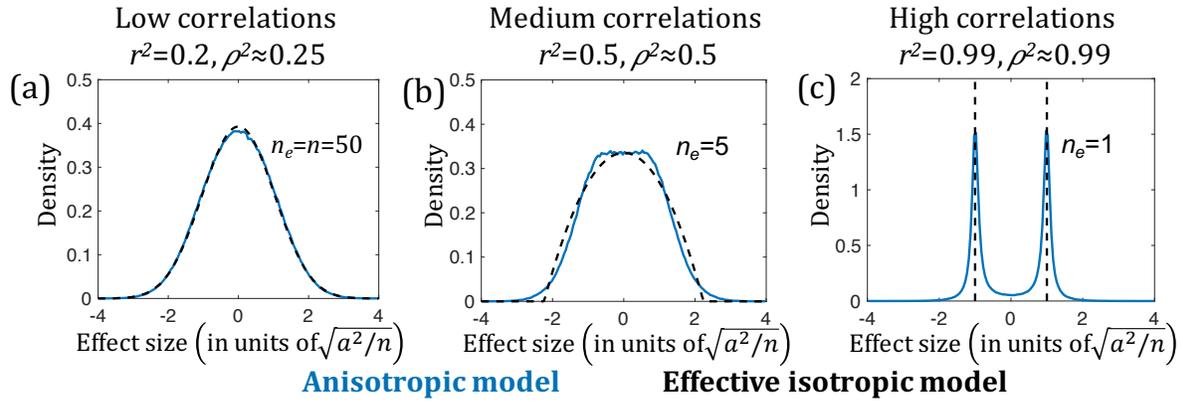

**Figure S10.** Effects of correlations among traits on the distribution of effect sizes. Numerical results for our model with the correlation matrix defined in Eq. S77 and $n = 50$ traits are shown in blue and the corresponding isotropic model in black dashes. (a) When correlations are low, the isotropic model approximates the distribution of effect sizes well. (b) With large correlations, we need to use an effective number of traits, in this example $n_e = 5$, and rescale selection, in this case to $w_e^2 = w^2 n_e/n = w^2/10$, in order to approximate the distribution of effect sizes. (c) When the correlations approach 1, the distribution of effect sizes becomes singular and approaches the distribution for an isotropic model with $n_e = 1$ and $w_e^2 = w^2/n = w^2/50$.

## 6. The power to detect loci in GWAS

In this section, we summarize the results that we rely on in connecting our theoretical results with the observations in GWAS (see Discussion in main text). These results provide a first approximation to the power to detect loci in GWAS in re-sequencing and genotyping studies. They neglect some potential complications, which lie beyond the scope of this study (e.g., synthetic associations (24, 25)).

### 6.1. Re-sequencing studies

First, we consider how the power to identify a locus in a GWAS depends on its contribution to genetic variance. To this end, we follow Sham and Purcell (24) in assuming a simplified model for a GWAS in which loci are detected using a linear regression of the phenotype against the genotype at individual loci, and the dependence of phenotype on genotype follows an additive model. The slope of the regression (the regression coefficient), which is also our estimate of the effect size, $\hat{a}_1$ is then approximately normally distributed as

$$\hat{a}_1 \sim N\left(a_1, \frac{V_P/m}{2x(1-x)}\right), \tag{S78}$$

where $a_1$ is the true effect size and $x$ is the minor allele frequency at the locus (which, due to the large study sizes, we assume is estimated without error), $V_P$ is the total phenotypic variance, and $m$ the study size (which in reality may be an effective size reflecting study design, e.g., when the sample was split into discovery and validation panels) (24).

Under the null hypothesis, the true effect size is 0, meaning that

$$\hat{a}_{1\,\text{null}} \sim N\left(0, \frac{V_P/m}{2x(1-x)}\right) \tag{S79}$$

and therefore, the estimated contribution to variance has a chi-squared distribution with one degree of freedom

$$\frac{\hat{v}_{\text{null}}}{V_P/m} = \frac{2\hat{a}_{1\,\text{null}}^2 x(1-x)}{V_P/m} \sim \chi_1^2. \tag{S80}$$

The power to identify a locus as significant with p-value $p^*$ is the probability that the estimated contribution of the locus to variance, $\hat{v}$, is large enough that

$$\Pr(\hat{v}_{\text{null}} > \hat{v}) < p^*. \tag{S81}$$

This condition can be translated into a threshold contribution to variance $v^*$ for which loci with $\hat{v} > v^*$ are considered significant, i.e. $\Pr(\hat{v}_{\text{null}} > v^*) = p^*$, with $v^*$ given by

$$\frac{v^*}{V_P/m} = 2\left(\text{erf}^{-1}(1-p^*)\right)^2, \tag{S82}$$

and erf denoting the error function. The power to identify a locus as significant would then be $\Pr(\hat{v} > v^*)$, and the distribution of $\hat{v}$ given by

$$\frac{\hat{v}}{V_P/m} = \frac{2\hat{a}^2 x(1-x)}{V_P/m} \sim \chi_1^2 \left(\frac{v}{V_P/m}\right), \quad (S83)$$

with $\chi_1^2$ denoting a non-central chi-squared distribution with one degree of freedom. Therefore, power is given by

$$H(v, p^*) = \Pr(\hat{v} > v^*) = h_+\left(\frac{v}{V_P/m}, p^*\right) + h_-\left(\frac{v}{V_P/m}, p^*\right), \quad (S84)$$

with $h_\pm(y, p^*) = \frac{1}{2}\left(1 \pm \mathrm{erf}\left(\sqrt{y/2} \mp \mathrm{erf}^{-1}(1-p^*)\right)\right)$ and the two terms correspond to the estimated and true effect sizes having the same or opposite sign.

The form of the power function carries important implications (Eq. S84 and Fig. S11). Notably, it shows that (in this approximation) power depends only on the contribution of a locus to variance, and this contribution should be measured relative to, or in units of, $V_P/m$. This scale makes intuitive sense, because the total phenotypic variance generates the background noise for detecting any individual locus, and the background noise is inverse proportional to the study size. In particular, the threshold contribution to variance $v^*$, as defined above, is proportional to $V_P/m$ and is also the contribution to variance at which power is 50%, i.e.,

$$H(v^*, p^*) = 1/2. \quad (S85)$$

The power function can then be approximated by a step function (see Fig. S11)

$$H(v) \approx \Theta(v - v^*) = \begin{cases} 1 & v > v^* \\ 0 & v < v^* \end{cases}. \quad (S86)$$

This will be a good approximation when the number of loci that fall at intermediate range (e.g., with power between 0.1 and 0.9) is negligible compared to the number that falls outside this range.

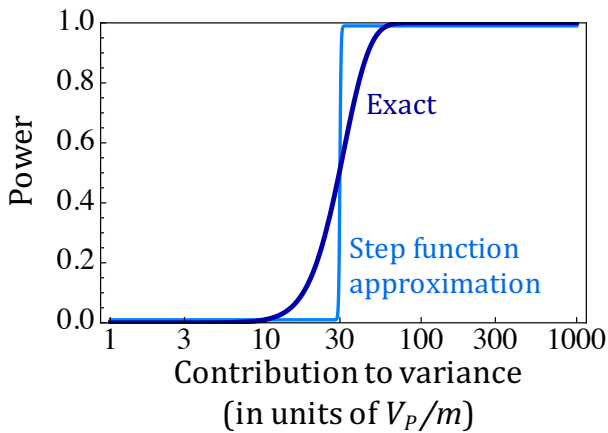

**Figure S11.** The power to detect loci as a function of their contribution to genetic variance (given in units of $V_P/m$). Shown are the exact power function (Eq. S84) and its step function approximation (Eq. S86) for $p = 5 \cdot 10^{-8}$.

Further insights come from considering this power function in conjunction with our theoretical results (Section S3). Notably, our results suggest that the first loci to be detected, those that contribute the most to variance, are weakly and strongly selected, and that their contributions to variance are on the scale of $v_s$. We therefore expect GWAS to begin to identify loci (and account for genetic variance) when the study size is such that $v^* \propto V_P/m$ is on the order of $v_s$, i.e., when $m \sim V_P/v_s$. We would further expect the rate of increase in identifying new loci (and accounting for the variance) to be similar for different traits when the variance is measured in units of $v_s$.

## 6.2. Genotyping

Most current GWAS rely on genotyping instead of re-sequencing, resulting in an additional loss of power (26). Specifically, these studies impute the alleles at loci that are not included in the genotyping platform (27), and the imputation becomes imprecise when the imputed alleles are rare (Fig. S12). If causal loci with rare alleles are included in GWAS, this imprecision leads to an under-estimation of their effect size, resulting in a loss of power (26). For loci with MAF $x$ and effect size $a$, the expected estimate of the effect size would be reduced by a factor of $r(x)$, where $r^2(x)$ is the mean correlation between the imputed and real alleles (28), and the distribution of estimates can be approximated by

$$\hat{a}_1 \sim N\left(ra_1, \frac{V_P/m}{2x(1-x)}\right). \tag{S87}$$

Employing the reasoning of the previous subsection, we can therefore approximate the power to detect a locus by $H(r^2v, p^*)$, where H is the power function defined in Eq. S84.

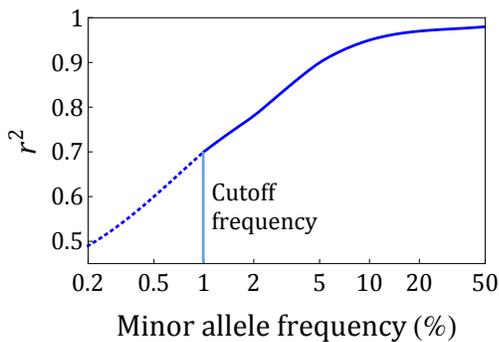

**Figure S12**. The precision of imputation decreases with MAF. Specifically we show the mean correlation between imputed and real genotypes as function of minor allele frequency, for a study using an Illumina 1M SNP array and the 1000 genomes phase III as an imputation panel (based on Extended Fig. 9A in (29)). We approximate the effect on power by excluding loci with MAF < 1% and assuming that loci with greater MAFs are imputed correctly.

In practice, GWAS often include only loci with MAF above a threshold, which is chosen to ensure precise imputation. We therefore approximate the effect of genotyping on power by excluding loci below a threshold MAF and assume that loci that exceed this threshold are imputed correctly.

### 6.3. Tagging in GWAS

Our inference is predicated on the assumption that the distribution of estimated variances among genome-wide significant (GWS) associations faithfully reflects the distribution among causal loci. We have no obvious alternative but to make this assumption, and arguably the good fit of our theoretical predictions to the distribution of variances among associations provides some support for this assumption. While it cannot be directly tested at present, existing arguments and evidence suggest that it is plausible, for reasons we review.

Most of the variants discovered by GWAS are common. Specifically, all but one of the GWS associations for height and BMI, which we rely upon in our inference, have MAF>1%, and the MAF of most associations is considerably greater. In considering the validity of our assumption, we therefore consider what could be tagged by such common associations. One possibility is that a given common association is tagging a single common, causal variant. Given the accuracy of imputation for common variants (see Fig. S12), we would therefore expect that the tagging variant would be in almost perfect LD with the causal one (including the possibility that the association is actually with the causal variant). If that were the case, then we would expect the estimated frequency and effect size, and thus the estimated contribution to genetic variance, to be very similar to those of the causal variant. A second possibility is that a given association tags several common causal variants within the same genomic region. The number of causal variants would likely be small, as otherwise the tagging allele is highly unlikely to be in LD with causal alleles that affect the trait in the same direction. If that were the case, given the accuracy of imputation of the causal alleles, we would expect conditional analysis (e.g., 30) to successfully distinguish between the different causal variants, thus returning us to the previous scenario.

A third possibility involves a common association tagging rare, causal variants (25). While a single rare, causal variant would have to have an unreasonably large effect size in order to result in a common GWS association (31), it has been argued that several rare, causal variants in the same genomic region may be tagged by a single "synthetic association" (25). In this case, the relatively low LD between the association and each of the causal variants would imply that the estimated contribution to variance of the association would have to be much smaller than the combined contribution of the causal variants (25, 31). If this were the case for many associations identified in GWAS, it would likely violate the premise of our inference.

However, multiple lines of evidence suggest that it is not a common occurrence. One is that, where data is available, associations often replicate across populations. For example, there is considerable overlap between GWS associations for height in Europeans and East-Asians (32). While we would not expect perfect replication even if associations were tagging single, common, causal variant, we would expect practically none if they were synthetic, both because the underlying rare, causal alleles would be less likely to be shared among populations and because the particular LD configuration that allows for their tagging in one population would likely break down in others (33, 34). A second is that simulation studies suggest that synthetic associations are expected to have much lower MAF than typically observed among associations in GWAS (31). Moreover, these simulations suggest that, because synthetic association should capture only a fraction of the variance contributed by the tagged loci, having many synthetic associations would imply there being much more heritable variance than is known to be present in the population. A third, and perhaps most direct line of evidence, is that, to the best of our knowledge, none of the studies that pursued fine-mapping around GWS associations have uncovered such synthetic associations (33, 35, 36). These arguments, together with other lines of evidence (e.g., 31) suggest that in practice synthetic associations are likely to be rare.

Perhaps a more plausible alternative is for an association to primarily tag one common, causal variant, with which it is in high LD, but also to pick up the effects of one or a few rare, causal variants, which are more poorly tagged. Under this scenario, we might expect the estimated contribution to variance to slightly overestimate the contribution of the dominant causal variant. To the best of our knowledge, this scenario has not been well characterized, making it difficult to assess how common it is or whether the overestimation would be substantial.

In summary, given what we now know, our assumption about the distribution of estimated variances among associations reflecting the distribution among causal loci seems sensible.

## 7. Inference

In this section, we describe how we used our model to make inferences based on GWAS results for height and body mass index (BMI). As we note in the Discussion, these inferences are meant as an illustration and do not incorporate the effects of demography and a few other factors (e.g., genotyping and errors in the estimation of effect sizes (24, 26)), which lie beyond the scope of this study.

### 7.1. The composite likelihood

Our inferences are based on a composite-likelihood approach. We begin by describing the composite-likelihood function and its maximization, when the loci detected by GWAS are strongly selected and can be described by the high-pleiotropy limit. In this case, we have shown that the distribution of variances among loci is insensitive to the distribution of selection coefficients, depends on a single parameter $v_s$, and is well approximated by the probability density

$$\rho(v) = \frac{2\exp(-2\sqrt{v/v_s})}{v} \tag{S88}$$

(Section S3.2). Further approximating the power in GWAS as a step function (see Section S6), we find that the probability density of sites that exceed a threshold $v^*$ can be approximated by

$$f(v|v_s, v^*) = \frac{\rho(v)}{\int_{v>v^*} \rho(v)} = \frac{\exp(-2\sqrt{v/v_s})}{2v\, \mathrm{I}(2\sqrt{v^*/v_s})}, \tag{S89}$$

where $\mathrm{I}(x) \equiv \int_{t>x} \exp(-t)/t$ (see Eq. S37). We therefore approximate the log-composite-likelihood of $v_s$ given the contributions to variance of the $K$ loci detected in a GWAS, $\{v_i\}_{i=1}^K$, by

$$\mathrm{LCL}(v_s|\{v_i\}_{i=1}^K, v^*) = \sum_{i=1}^K \log(f(v_i|v_s)) =$$

$$= -(2/\sqrt{v_s})\sum_{i=1}^K \sqrt{v_i} - K\log\left(\mathrm{I}(2\sqrt{v^*/v_s})\right) - \sum_{i=1}^K \log(\sqrt{v_i}). \tag{S90}$$

It follows that the composite-likelihood is maximized when

$$\hat{v}_s = \mathrm{argmin}_{v_s}\left\{2\overline{\sqrt{v}}/\sqrt{v_s} + \log\left(\mathrm{I}(2\sqrt{v^*/v_s})\right)\right\}, \tag{S91}$$

where $\overline{\sqrt{v}} \equiv \frac{1}{K}\sum_{i=1}^K \sqrt{v_i}$.

We also consider the models without pleiotropy and in which the degree of pleiotropy is a parameter. In the case without pleiotropy,

$$\rho(v) = \frac{2\exp(-2v/v_s)}{v} \tag{S92}$$

(see Section S3.2). By following the same steps, we find that the composite-likelihood is then maximized when

$$\hat{v}_s = \mathrm{argmin}_{v_s}\{2\bar{v}/v_s + \log(\mathrm{I}(2\,v^*/v_s))\}, \tag{S93}$$

where $\bar{v} \equiv \frac{1}{K}\sum_{i=1}^{K} v_i$. When the degree of pleiotropy $n$ is a parameter of the model, we find that

$$\rho_n(v) = \int_{a_1} \frac{2}{v} \exp\left(-\frac{2v/v_s}{a_1^2/(a^2/n)}\right)\varphi_n(a_1|a) \tag{S94}$$

(see Section S3.2). Again, following the same steps, we find that the probability density of sites that exceed a threshold $v^*$ is

$$f_n(v|v_s) = \frac{\rho_n(v)}{\int_{v>v^*}\rho_n(v)} \tag{S95}$$

and the log-composite-likelihood is

$$\mathrm{LCL}(v_s, n|\{v_i\}_{i=1}^{K}, v^*) = \sum_{i=1}^{K}\log(\rho_n(v_i)) - K\log\left(\int_{v>v^*}\rho_n(v)\right). \tag{S96}$$

In the latter case, we used numerical maximization to show that the composite-likelihood estimates for height and BMI converge to the high pleiotropy limit. Specifically, we maximized the composite-likelihood specifying an interval of [1,1000] for $n$, where for both traits the estimates converged to the upper limit of 1000. While numerical optimization does not allow us to specify an infinite interval, the likelihood function and maximal value for $n=1000$ are indistinguishable from those in the high-pleiotropy limit.

## 7.2. Determining $v^*$ and removing outliers

Our likelihood maximization requires us to specify the value of the threshold $v^*$. We choose this threshold based on the empirical distributions of the contributions to variance among genome-wide significant associations (Fig. S13A & B). Specifically, when the contributions to variance approach the lower boundary for discovery, we observe a decline in the density of loci. This is likely due to a gradual reduction in power and suggests that our approximation for power (as a step function) breaks down for these values of variance. We therefore choose thresholds that appear to be above this decline ($v^* = 1.4 \cdot 10^{-4}V_P$ for height and $v^* = 1.35 \cdot 10^{-4}V_P$ for BMI; Fig. S13A & B), resulting in the removal of 53 loci for height and 11 for BMI. We also examine how our estimates of $v_s$ depend on the choice of $v^*$, and find that they are much more sensitive to reducing the threshold than to increasing it; in fact, the estimates we obtain by increasing the threshold are within the confidence intervals of the estimate with the chosen thresholds (Fig. S13C & D). This analysis further supports our choice to exclude the loci with the lowest contribution to variance. For BMI, we also dropped the locus with the largest contribution to variance (near FTO), which appears to be an outlier (Fig. S13B) and has been suggested to be under balancing selection (37).

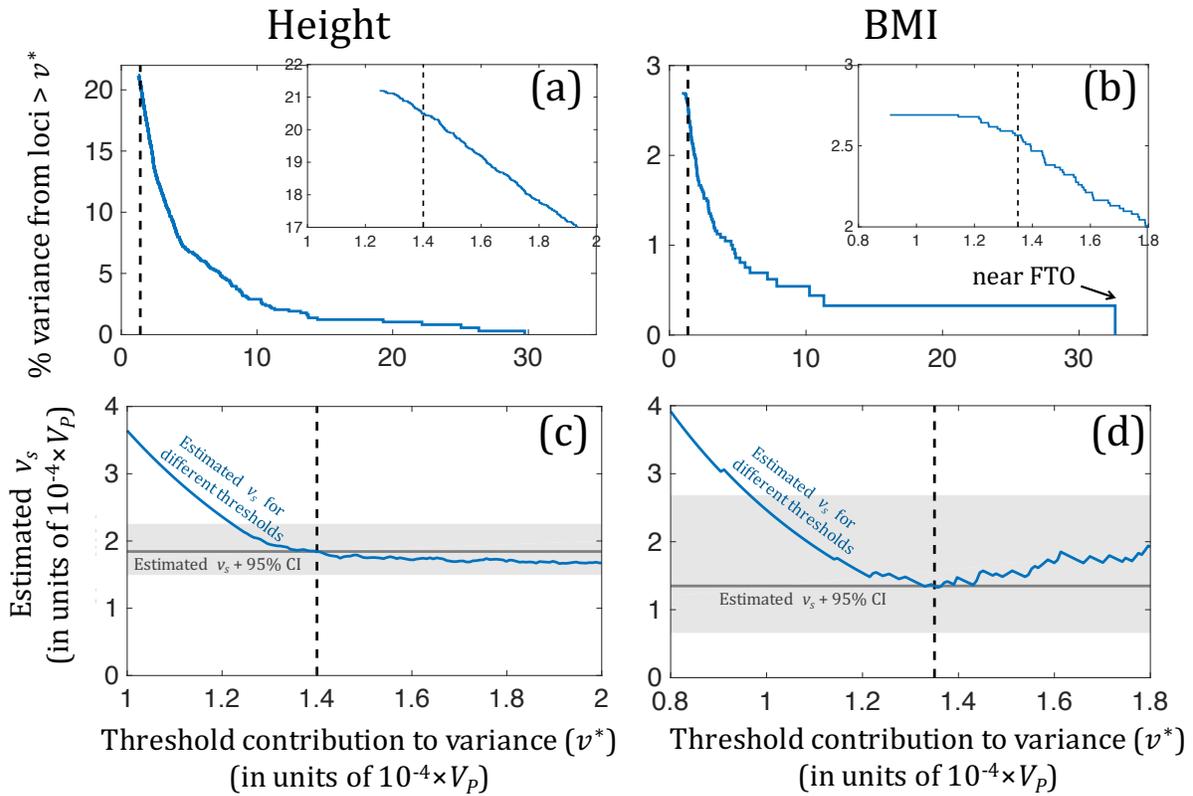

**Figure S13.** Determining $v^*$ and removing outliers. The total variance from significant associations as a function of the threshold contribution to variance, for height (a) and BMI (b). The insets show a close up of the lower range of contributions to variance, highlighting the decline in the density of discovered loci. Our chosen thresholds are shown by the dashed vertical line (in all graphs). Our estimates of $v_s$ as a function of the chosen threshold, for height (c) and BMI (d). When we increase the threshold, the estimates remain within the 95% CI of the estimate with our chosen threshold.

### 7.3. Estimating target size and explained variance

We estimate the target size and the variance explained, both for varying study size and total, based on our estimates of $v_s$. The population-scaled mutational input per generation from strongly selected loci, $2NU_s$, is estimated by

$$\widehat{2NU_s} = K/\int_{v>v^*} \rho(v|\hat{v}_s), \tag{S97}$$

(see Eq. S38) and the corresponding estimate for the target size is

$$\hat{L}_s = \widehat{2NU_s}/\widehat{2Nu}, \tag{S98}$$

where the estimate for the population scaled mutation rate per site per generation $\widehat{2Nu} \approx 0.5 \cdot 10^{-3}$ is based on heterozygosity (29). The explained variance corresponding to GWAS with study size $m$ is estimated by

$$\hat{\sigma}_s^2(m) = \widehat{2NU_s} \int_{v>v^*(m)} v\rho(v|\hat{v}_s) = K\left(\int_{v>v^*(m)} v\rho(v|\hat{v}_s)/\int_{v>v^*(m_0)} \rho(v|\hat{v}_s)\right), \tag{S99}$$

where we approximate the threshold corresponding to study size $m$ based on the study size, $m_0$, and threshold, $v^*$, in current GWAS, by

$$v^*(m) = v^* \cdot (m_0/m). \tag{S100}$$

To estimate the total variance arising from strongly selected loci, we simply set the threshold in Eq. S99 to 0.

### 7.4. Estimating confidence intervals

We use a combination of non-parametric and parametric bootstrap to estimate confidence intervals (CI). We use non-parametric bootstrap to estimate the CI for the model parameters $v_s$ and $L_s$: specifically, we perform 10,000 iterations, in which we resample the loci identified by GWAS and repeat the estimation of $v_s$. We use parametric bootstrap to estimate the confidence intervals in Fig. 5A, describing the explained variance as a function of threshold based on our model. To that end, we rely on our model with the point estimates for $v_s$ and $L_S$, to generate 10,000 samples from GWAS with the specified threshold, and then calculate the total variance explained by these samples. We use a combination of non-parametric and parametric bootstrap to calculate the CI for model predictions, including the total variance, $\sigma_s^2$, and the explained variance, $\sigma_s^2(m)$, and number of loci as a function of study size (Fig. 5B & C). In this case, we generate 10,000 samples by: i) estimating $v_s$ based on a resampled set of GWAS loci (similar to the non-parametric procedure), and ii) using the estimated $v_s$ and corresponding $L_s$ to generate a GWAS hits above $v^*$ based on our model (similar to the parametric procedure); we then calculate the appropriate summary based on the latter samples. This two stage procedure is intended to capture the uncertainty generated by both the errors in estimating our basic model parameters and the noise generated by the stochastic processes underlying the number and variance at segregating loci that are yet to be discovered. The resulting estimates and CI are summarized in Table S2.

| Parameter | | Height | BMI |
|---|---|---|---|
| Contribution to variance per strongly selected locus (in units of the total phenotypic variance) | $\hat{v}_s/V_P$ | 1.8 [1.5, 2.3]×10$^{-4}$ | 1 [0.6, 1.7] ×10$^{-4}$ |
| Expected study size required to describe 50% of the strongly selected variance | $m_{50\%}$ ($\approx 43 V_P/\hat{v}_s$) | 230 [190, 290] K | 420 [250, 770] K |
| Number of newly arising strongly selected mutations per generation in the population | $2NU_s$ | 2300 [1800, 3000] | 600 [300, 1900] |
| Mutational target size for strongly selected mutations | $L_s$ | 4.6 [3.6, 6.0] Mbp | 1.3 [0.6, 3.8] Mbp |
| % contribution to phenotypic variance from strongly selected loci | $\sigma_s^2/V_P$ | 42 [39, 45] % | 7 [5, 10] % |
| Proportion of heritability from strongly selected loci | $\sigma_s^2/V_G$ ($= \sigma_s^2/h^2 V_P$) | 53 [49, 57] % | 13 [10, 21] % |

**Table S2.** Parameter estimates and their confidence intervals for height and BMI based on GWAS results; the heritability was assumed to be 0.8 for height and 0.5 for BMI (8, 10).

### 7.5. Testing goodness of fit

We use the Kolmogorov-Smirnov D statistic (38, 39) to test the goodness of fit of our models without pleiotropy and in the high pleiotropy limit. Since our parameter estimates are inferred from the data that we are testing against, we cannot rely on the standard tables for the p-values. We therefore generate null distributions for the D statistic using parametric bootstrap based on our models. Specifically: i) we generate $10^5$ samples of $K$ significant loci based on the model under consideration, with the corresponding estimate of $v_s$, ii) we infer $v_s$ based on each sample, and iii) calculate the Kolmogorov-Smirnov D statistic between the distribution of variances for the $K$ loci in each sample and the corresponding theoretical distribution based on the $v_s$ inferred from that sample. The resulting distribution of D statistics corresponds to our null hypothesis, i.e., that the loci detected in GWAS arose according to our model, and specifically to the way we calculate the D statistic between the observed distribution of variances for the K detected loci and the theoretical distribution that we inferred based on these observations. We then calculate the D statistic, $D_r$, based on the real data and corresponding theoretical distribution, and estimate the one-sided p-value by

$$\hat{p}_{K-S} = \frac{\text{\# simulated datasets with } D > D_r}{\text{\# simulated datasets}}. \tag{S101}$$

Note that unlike the common case, here the inability to reject the null indicates that the data is consistent with our model.

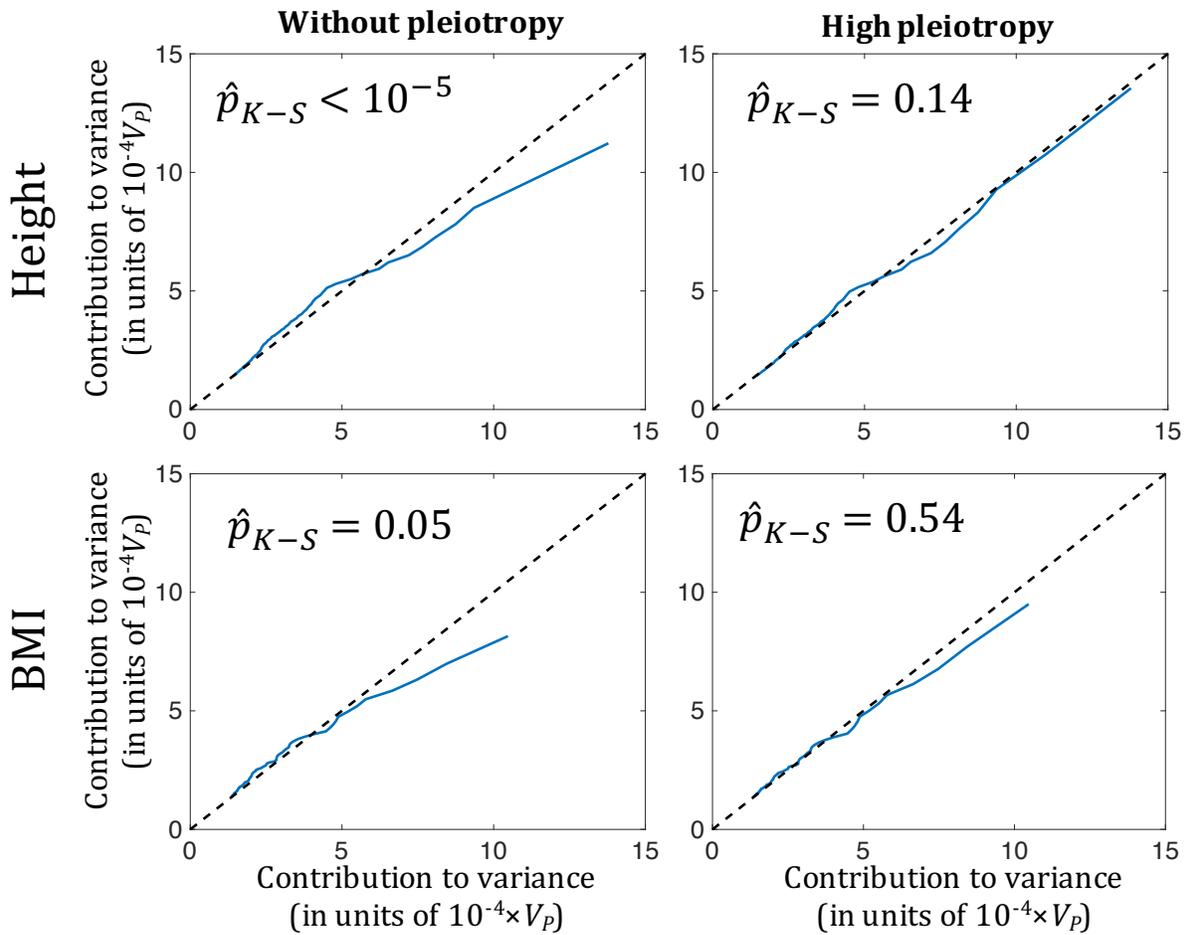

Figure S14. Q-Q plots comparing the distribution of variances among significant loci taken from the GWAS of height (10) and BMI (8) with the theoretical distributions inferred from these data, based on the models without pleiotropy (a) and in the high pleiotropy limit (b). These plots show that the model assuming high pleiotropy cannot be rejected for either trait and fits these data much better than the model without pleiotropy.

## 8. Consistency with other datasets and analyses

Here, we show that the results of our inference for height are consistent with findings of a recent GWAS based on exome genotyping; that our inferences for height and BMI are consistent with estimates of the heritability tagged by SNPs with MAF > 1% in the GWAS we used; and that our model is consistent with estimates about the relationship between effect size and MAF in these and other GWAS.

### 8.1. Exome association study of height

Marouli et al. (40) present an association study for height that was specifically designed to capture rare, exonic variants. They rely on the ExomeChip genotyping array (41), which includes the vast majority of protein-altering variants with MAF>0.1%, allowing them to directly (i.e., without imputation) test for associations among rare variants. Using a study size of more than 300,000 European individuals, they find over 400 genome-wide significant associations. Here we examine whether their findings are consistent with our inference based on the Wood et al. genome-wide, genotyping based GWAS for height (10).

In addition to protein altering variants, the ExomeChip includes some synonymous SNPs and ancestry informative markers, as well as all of the genome-wide significant associations listed in NHGRI from 2011. To avoid ascertainment biases, we consider only protein-altering variants, including non-synonymous, splice region, splice acceptor and stop codon variants. This leaves us with 250 of the Marouli et al. genome-wide significant associations. In addition, we apply the procedure described in Section S7.2, resulting in the removal of associations with contributions to variance below $v_E^* = 1.15 \cdot 10^{-4} \, V_P$, for which power is substantially diminished (Fig. S15A) ; this step leaves us with 147 associations. Next, we compare the distribution of variances among the remaining 147 associations with our theoretical prediction, with the $v_S$ inferred from the Wood et al. data (Table S2) above the threshold $v_E^*$ (Fig. S15B). We do not consider the fit to the number of associations, because it depends on the mutational target size for protein-altering variants affecting height, which is unknown.

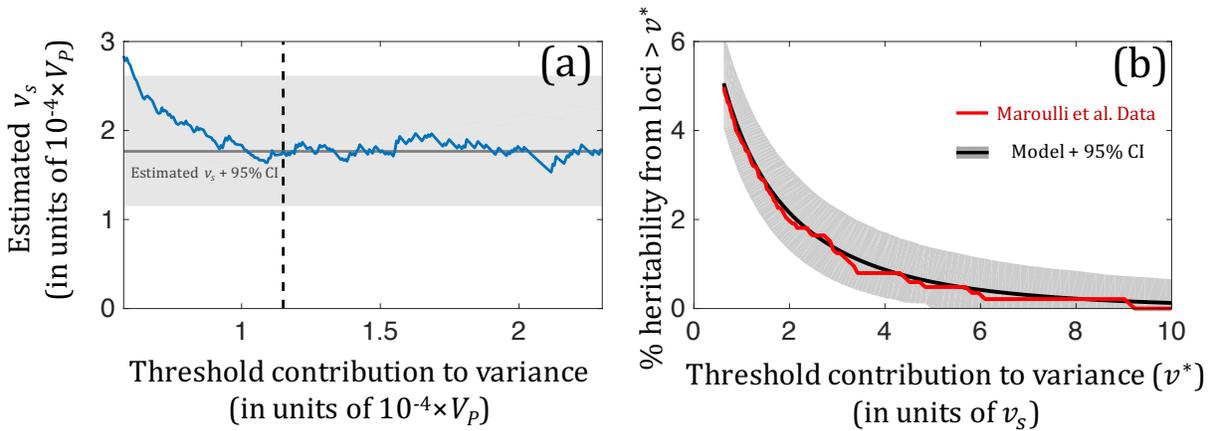

Figure S15. Comparing our inferences for height with the results of the Marouli et al. GWAS. (a) Choosing the threshold contribution to variance, $v_E^*$, above which our approximation for power applies; see Section S7.2 for details. (b) Comparing the predicted and observed distribution of variances above the threshold $v_E^*$. 95% CIs for our predictions are based on bootstrap; see Section S7.4 for details.

To test whether the observed distribution is consistent with our prediction, we calculate the Kolmogorov-Smirnov $D$ statistic (38, 39) for this comparison, $D_r$, and ask whether we can reject our prediction based on the value of $D_r$. In approximating the null distribution of the $D$ statistic, we must consider that: i) Some of the Marouli et al. associations might have been tagged by the genome-wide significant associations in Wood et al., which we relied upon in estimating $v_S$; this would lead to smaller values of the $D$ statistic than if the two sets of associations were independent. ii) Our estimate of $v_S$ includes some statistical error, due to the finite set of associations on which it relies. To account for these factors, we employ a parametric bootstrap procedure that mimics how the value of the $D$ statistic arises, under the conservative scenario in which any of the associations from Marouli et al. could have been included in the data that we used in our inference. Specifically, we assume that the distribution of variances among loci follows the theoretical prediction with our estimate of $v_S$, and i) We sample 147 associations from the predicted distribution with threshold $v_E^*$, corresponding to the Marouli et al. associations. ii) Given the number, $k$, of these associations that fall above the threshold of the Wood et al. GWAS, $v_G^* = 1.4 \cdot 10^{-4} V_P$ (Section S7.2), we sample an additional $644 - k$ variants from the predicted distribution with threshold $v_G^*$. The resulting 644 simulated associations that fall above $v_G^*$ correspond to the Wood et al. associations. iii) We infer $\hat{v}_S$ based on these 644 variants, thus mimicking our inference procedure, and calculate the $D$ statistic for our predicted distribution with $\hat{v}_S$ and the distribution based on the 147 simulated variants. iv) We repeat this procedure $10^5$ times to approximate the distribution of $D$ statistic under our null, and estimate the one-sided p-value by

$$\hat{p}_{K-S} = \frac{\text{\# simulated datasets with } D>D_r}{\text{\# simulated datasets}}. \tag{S102}$$

Doing so, we find that $\hat{p}_{K-S} = 0.99$, and thus, we cannot reject our predictions based on the data from Marouli et al. (40). This result indicates a good fit to their findings.

## 8.2. The heritability arising from common SNPs

Yang et al. (42, 43) estimate the heritability that is tagged by common SNPs (MAF>1%) in GWAS of several traits, including height and BMI. Here we ask whether their estimates are consistent with our inferences based on genome-wide significant (GWS) associations from the same GWAS.

First, we consider our inferences predicated on equilibrium demography. On this assumption, we predict that GWS associations would be under intermediate or strong selection, roughly corresponding to $S > 5$. Our estimates then suggest what proportion of variance arises from loci under this range of selection effects, where the rest of the variance is assumed to arise from loci under weaker selection. The proportion of variance that arises from sites with $S < 5$ and MAF $> 1\%$, $P_w(> 1\%)$, can be bound from above by the variance that would arise if they were all effectively neutral, $P_n(> 1\%)$. Further denoting the proportion of variance that arises from sites with $S > 5$ and MAF $> 1\%$ by $P_s(> 1\%)$, and the overall proportion of variance from sites with MAF $> 1\%$ by $P(> 1\%)$, we obtain the following requirement:

$$P_s(> 1\%) = P(> 1\%) - P_w(> 1\%) \geq P(> 1\%) - P_n(> 1\%). \tag{S103}$$

For height, Yang et al. estimate that $P(> 1\%) = 0.59$ (42), and our estimate for $P_n(> 1\%) = 0.45$. As Fig. S16 shows, so long as most of the estimated variance with $S > 5$ (53%) arises from loci with $S < 135$, the requirement in Eq. S103 will be easily met. For BMI, Yang et al. estimate that $P(> 1\%) = 0.5$ (42), and our estimate for $P_n(> 1\%) = 0.83$. The lower bound in Eq. S103 is therefore negative, implying that requirement S103 is met regardless of the distribution of selection coefficients for $S > 5$.

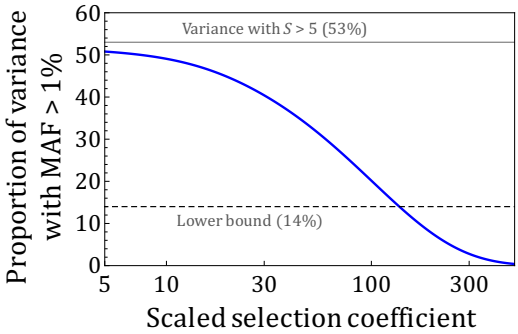

**Figure S16.** The Yang et al. (42, 43) estimate of the genetic variance in height arising from loci with $MAF > 1\%$ imposes weak constraints on the distribution of selection coefficients, assuming our estimate for the genetic variance with $S > 5$.

Next, we consider the results of our analysis in Section 9, incorporating the effects of recent changes in European populations size. Our results suggest that GWS associations arise from loci with selection coefficients of $s \approx 10^{-3}$. We therefore ask whether the Yang et al. (42, 43) estimates are consistent with ours, when we attribute our equilibrium estimates of the proportion of variance arising from intermediate and strongly selected loci to selection coefficients of $s \approx 10^{-3}$, assuming that the remaining variance arises from loci under weaker or stronger selection (a more rigorous approach would be to account for demography in estimating the proportion of variance, but this extension lies beyond the scope of the current paper). The proportion of variance arising from sites under weaker selection with MAF $> 1\%$ is bound from above by $P_n(> 1\%)$, whereas the corresponding proportion from sites under stronger selection can be vanishingly small. Denoting the proportion of variance arising from sites with $s \approx 10^{-3}$ and MAF $> 1\%$ by $P_{10^{-3}}(> 1\%)$, we therefore obtain the following condition:

$$P(> 1\%) \geq P_{10^{-3}}(> 1\%) \geq P(> 1\%) - P_n(> 1\%). \tag{S104}$$

If we assume the Yang et al. (42) estimates for $P(> 1\%)$ and our estimates for $P_{10^{-3}}(> 1\%)$, Table S3 shows that this requirement is easily met for both height and BMI. More generally, our analysis illustrates that heritability estimates of this kind impose rather weak constraints on our inferences.

|  | $P(> 1\%)$ |  | $P_{10^{-3}}(> 1\%)$ |  | $P(> 1\%) - P_n(> 1\%)$ |
| --- | --- | --- | --- | --- | --- |
| Height | 0.59 | $\geq$ | 0.48 | $\geq$ | 0.59−0.38=0.21 |
| BMI | 0.5 | $\geq$ | 0.12 | $\geq$ | 0.5−0.83=−0.33 |

**Table S3.** Consistency between the Yang et al. (42) estimates of the total variance arising from loci with MAF $> 1\%$ and our estimates of the variance arising from sites with $s \approx 10^{-3}$ and MAF $> 1\%$.

### 8.3. The relationship between SNP heterozygosity and effect size

More recent studies of the heritability tagged by SNPs in GWAS also make inferences about the relationship between effect sizes and MAF (44, 45). Specifically, they assume that the relationship between the contribution of a site to variance, $v = 2a_1^2 x(1-x)$, and its MAF, $x$, takes the form

$$E(v|x) \propto \left(x(1-x)\right)^{\alpha+1}, \tag{S105}$$

or equivalently, that

$$E(a_1^2|x) \propto \left(x(1-x)\right)^{\alpha}, \tag{S106}$$

and they estimate the value of $\alpha$ from the data.

Provided a distribution of selection coefficients, f($S$), Eq. S20 implies that in our model

$$E(a_1^2|x) = \frac{\int_{a_1} a_1^2 \rho(x,a_1)}{\int_{a_1} \rho(x,a_1)} = \frac{\int_S \int_{a_1} a_1^2 f(S)\tau(x|S)\eta(a_1|S)}{\int_S \int_{a_1} f(S)\tau(x|S)\eta(a_1|S)}$$

$$= \frac{2w^2}{nN} \cdot \frac{\int_S S\, f(S)\tau(x|S)}{\int_S f(S)\tau(x|S)} = \frac{2w^2}{nN} \cdot E(S|x). \tag{S107}$$

Thus, in our model, assuming the relationship of Eq. S105 (or S106) would imply that

$$E(S|x) \propto \left(x(1-x)\right)^{\alpha}. \tag{S108}$$

The aforementioned studies assume the relationship in Eq. S105 (or S106), without providing any evidence that this somewhat arbitrary functional form fits the data better than others, and show that values of $\alpha$ between -1 and 0 provide the best fit to data from GWAS of a variety of traits. To show that our model is in agreement with theirs, all we therefore need to do is to find distributions of selection coefficients, f($S$), that approximate the relationship of Eq. S108 for values of $\alpha$ between -1 and 0. In Fig. S17, we assume that selection coefficients follow a Gamma distribution, where we vary its expectation and variance. As expected, $E(S|x)$ monotonically decreases as $x$ increases. When $E(S) \ll 1$ or the coefficient of variation $C_V^2(S) \ll 1$, $E(S|x)$ varies minimally with $x$ and can approximated by Eq. S108 with $\alpha = 0$. In other cases, $E(S|x)$ varies more substantially with $x$. When we approximate those cases using Eq. S108, we obtain a range of $\alpha$ values between $-1$ and 0. Thus, our model appears to be consistent with the values of $\alpha$ reported in (44) and (45). Our inferences for height and BMI are not very informative about the distribution of selection coefficients and are therefore not comparable with estimates of $\alpha$.

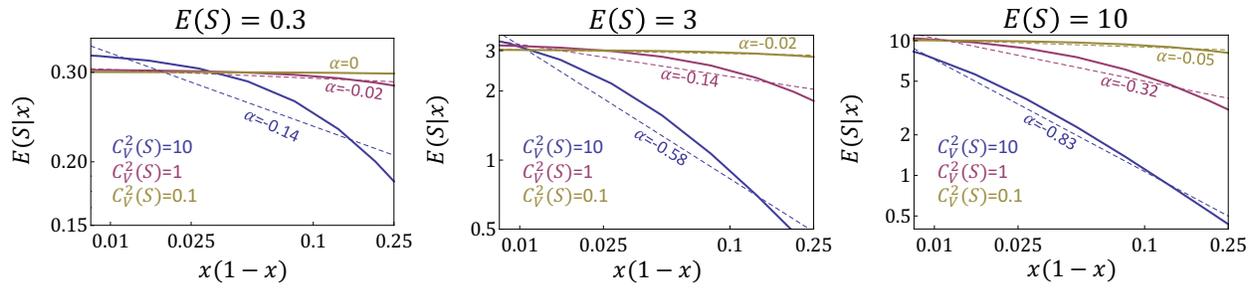

**Figure S17.** The relationship between effect size, or equivalently, selection coefficient, and MAF, shown on a log-log scale. Selection coefficients are gamma-distributed, with $E(S) = 0.3, 3, 10$ and shape parameters $k = 0.1, 1, 10$. $E(S|x)$ was approximated using the functional form $E(S|x) \propto (x(1-x))^\alpha$ (Eq. S108), by taking the values of $\log(E(S|x))$ and $\log(x(1-x))$ on a grid of $x$ values, $x = 0.5 \cdot 10^{-i/4}$ with $i = -8, -7, \ldots, 0$, and preforming least-square linear regression.

## 9. The effects of demographic history

While our theoretical results were derived on the assumption of a panmictic population of constant size, the evolutionary history of human populations sharply deviates from these simplifying assumptions. Notably, most large GWAS, including the studies of height (10) and BMI (8) that we use to test our predictions, have been performed in predominantly European populations, which are known to have experienced dramatic changes in their effective population size, including an Out-of-Africa bottleneck about ~100 KYA and explosive population growth over the past ~5 KY (46-49). The dramatic changes in population size have dramatically impacted the frequencies of neutral and selected alleles (46-48, 50-52), and are therefore expected to substantially impact the architecture of quantitative traits (51, 52). These considerations raise several questions about the interpretation of the fit between our predictions and GWAS data. Notably, how will these historical changes in population size affect our prediction, and specifically, why do our equilibrium predictions fit GWAS data despite the dramatic historical changes in population size? While a comprehensive treatment of these questions warrants a study in itself, we briefly address them here.

Even with changing population size, our results for the dynamics at segregating sites should still hold. Notably, we would expect the mean phenotype in the population to maintain the optimal phenotypic value, because any displacement from the optimum would be quickly adjusted by small changes to allele frequencies at numerous loci (see Section S4.4). As a result, the dynamics at individual sites would be decoupled, and well approximated by the first two moments of change in allele frequency described in Eqs. 5 and 6. In particular, the first moment would correspond to under-dominant selection, and the selection coefficient would be proportional to the size of the allele in the $n$-dimensional trait space (as described in Eq. 7). We can therefore study the effect of historical changes in population size on allele frequencies with simulations, using a fixed (not population-scaled) selection coefficient with under-dominance, and having the population size change over time.

To this end, we modify the simulation from Simons et al. (52) to incorporate under-dominance, and the historical changes in the effective population size of European populations inferred by Schiffels and Durbin (49) (Fig. S18). In brief, we simulate a bi-allelic site in a diploid, panmictic population, in which mutations, with selection coefficient $s$, arise at rate $u = 1.25 \cdot 10^{-8}$ per bp per generation (5, 49), and the next generation derives from Wright-Fisher sampling and fecundity selection. The simulation begins 150K generations ago (corresponding to 4.5 MYA with a generation time of 30 Y, as assumed by (49)), with a burn-in period with a constant population size of 14,448. In accordance with the Schiffels and Durbin inferences (49), changes in population size begin 55,940 generations ago (corresponding to 1.7 MYA). Specifically, we piece together the MSMC inferences from two and four haplotypes of European individuals (CEU) from HapMap project (53), where the four haplotype MSMC captures the bottleneck and recent growth and is used for times <170 KYA, and the two haplotype MSMC captures more ancient times and is used for times >170 kya (see Fig. S18). The derived allele frequency is recorded at the last generation corresponding to the present. The software and documentation can be found at https://github.com/sellalab/GenArchitecture.

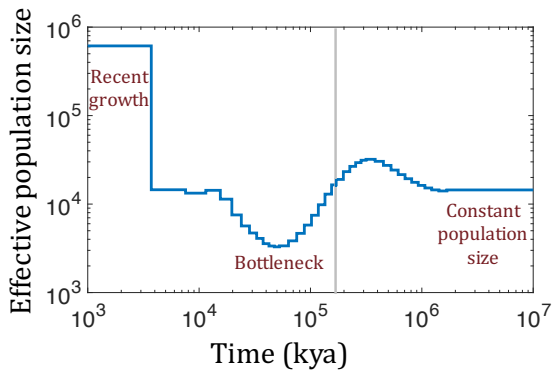

Figure S18. Changes in population size in the history of Europeans, as inferred by Schiffels and Durbin using MSMC (49). The cutoff between the two and four haplotype MSMC inferences is marked by the gray line.

We rely on such simulations to study how changes in populations size will affect the genetic architecture of a trait under the assumptions of our model. To this end, we consider a grid of selection coefficients: $s = 10^{-i/8}, i = 8, 9 \dots, 40$, where for each selection coefficient, we run $15 \cdot 10^6$ simulations. In this way, we obtain numerical approximations for the expected site frequency spectrum corresponding to each selection coefficient, which replaces the term $2Nu \cdot \tau(x|S)$ in our expressions for summaries of genetic architecture (Section S3). We further assume the high pleiotropy limit form for the distribution of effect sizes on the focal trait corresponding to a given selection coefficient (i.e., Eq. 11).

We first consider how demography affects the distribution of genetic variances among sites with different selection coefficients (Fig. S19A). The expected contribution per site (including both sites that are segregating and monomorphic) peaks around a selection coefficient of $s \approx 10^{-3}$ and, as in the case with constant population size (Fig. 2A), when the strength of selection increases, it appears to approach a plateau (Figs. S19A). The distribution of variances among sites, however, is dramatically affected by changes in population size: for selection coefficients around $s \approx 10^{-3}$, a much greater proportion of variance comes from sites with large contributions than from those with both weaker and stronger selection coefficients (Fig. S19B). This behavior contrasts with the case of a constant population size, where for sufficiently strong selection ($S > 5$), the distribution of variances among sites is insensitive to the strength of selection (see Fig. 3B).

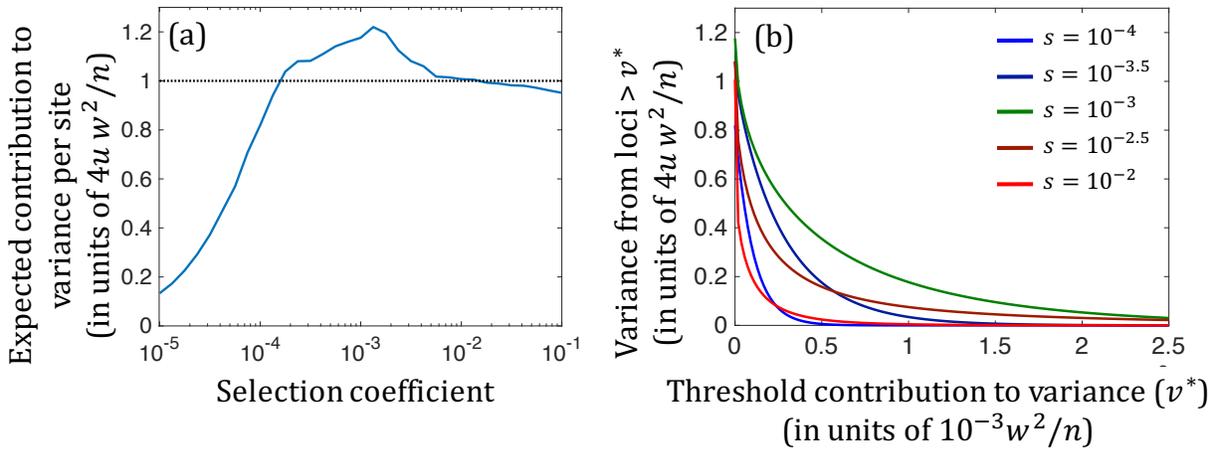

**Figure S19.** The joint effects of selection and changes in populations size (as inferred for Europeans by Schiffels and Durbin (49)) on the distribution of genetic variance among sites. (a) The expected contribution to variance per site, both segregating and monomorphic, as a function of the (unscaled) selection coefficient. Variance is measured in units of $4u\, w^2/n$, the equilibrium expectation for a strongly selected site. (b) The cumulative variance arising from sites with contributions above a threshold (y-axis) as a function of the threshold (x-axis); cumulative variance is measured in units of $4u\, w^2/n$, while the threshold in units of $10^{-3} w^2/n$.

As we establish below, these findings can be understood as follows. The segregating sites with the largest contribution to current genetic variance are due to mutations with $s \approx 10^{-3}$ that arose shortly before or during the Out-of-Africa bottleneck. Such mutations were under strong selection (i.e., with $2N_e s \approx 50$) before the bottleneck, but with the drop to an effective population size of $N_e \approx 4000$ during the bottleneck, they experienced more relaxed selection (with $2N_e s \approx 10$), allowing some of them to ascend to higher frequencies. The durations of subsequent increases in population size, and of explosive growth in particular, were too short to allow for a substantial reduction in their frequencies (e.g., a mutation with $s = 10^{-3}$ that reached 20% frequency by the end of the bottleneck, 15 Kya, would have an expected frequency of 18% at present). As a result, these mutations would have large contributions to variance at present. Moreover, their site frequency spectrum and distribution of contributions to variance are well approximated by assuming a population size of $N_e \approx 5000$ – roughly the geometric mean of populations sizes from the beginning of the bottleneck to the present – and thus to scaled selection coefficients of $2N_e s \approx 10$.

Extant segregating mutations under substantially stronger selection are expected to be much younger. They therefore tend to have arisen after the bottleneck, when the population size was considerably larger. As a result, they have much lower frequencies and per segregating site contributions to variance at present. The larger population size, however, will also increase the mutational input and thus the number of extant segregating sites; so long as selection is sufficiently strong, these effects balance each other such that the per site contribution to variance, counting both segregating and monomorphic sites, remains insensitive to changes in population size (52). In turn, extant segregating mutations under substantially weaker selection are expected to contribute much less variance per site (considering either segregating sites alone or all sites) primarily because of their smaller effect sizes, which is the same reason that applied in the case with a constant population size (see Fig. 2A).

We find support for this verbal argument when we relate the results of our simulations with the findings from GWAS. To do so, we follow the same reasoning that we applied to the case with constant population size (see Discussion). Namely, based on the distribution of variances (Fig. S19A), we would expect sites with selection coefficients around $s \approx 10^{-3}$ to be the first to be discovered in GWAS. Further assuming that such sites account for most associations discovered in GWAS and that their distribution of variances corresponds to $N_e = 5000$, we can use our estimates of $v_s$ for height and BMI to calculate the parameter $w^2/n$ ($= \frac{1}{2} N_e v_s$) for these traits. This approach allows us to plot the putative distribution of variances among sites that exceed the study thresholds, $v^*$, for different selection coefficients (Fig. S20A & B). Doing so, we find that the observed and fitted distributions are well approximated by the distributions for sites with $s \approx 10^{-3}$, thus supporting our premise that most of the explained variance arises from such sites, and that their distribution of variances is well approximated by assuming a constant population size of $N_e \approx 5000$. Our simulations also suggest that the proportion of variance explained for sites with $s \approx 10^{-3}$ is much greater than the proportion for sites under weaker or stronger selection (Fig. S20C & D), and should therefore also be greater than the total proportion of variance explained by these GWAS. This expectation accords with our findings as well, with our simulations suggesting that the proportion of variance explained for sites with $s \approx 10^{-3}$ is ~40% for height and ~30% for BMI (Fig. S20C & D) compared to a total proportion of ~25% for height and ~5% for BMI in these GWAS (8, 10).

Examining the expected MAF and allelic ages at sites that we predict to have been identified by these GWAS lends further support to our interpretation (Fig. S20E-H). Notably, we find that the MAF for sites with $s \approx 10^{-3}$ that are predicted to have been identified by these studies are similar to those that are observed (Fig. S20E & F). Moreover, when we examine the ages of mutations at detected sites, we find that mutations at sites with $s \approx 10^{-3}$ are predicted to have originated during or shortly before the OoA bottleneck (Fig. S20G & H).

In summary, our analyses suggest that the bulk of associations identified in the GWAS for height and BMI tag segregating mutations with $s \approx 10^{-3}$, which originated shortly before or during the OoA bottleneck. As a result, we would expect the distribution of variances among these sites to be well approximated by our equilibrium predictions corresponding to an effective population of $N_e \approx 5000$. This finding provides an explanation for why our equilibrium predictions fit the findings of GWAS in Europeans, despite our ignoring the dramatic changes in population size during their recent evolutionary past.

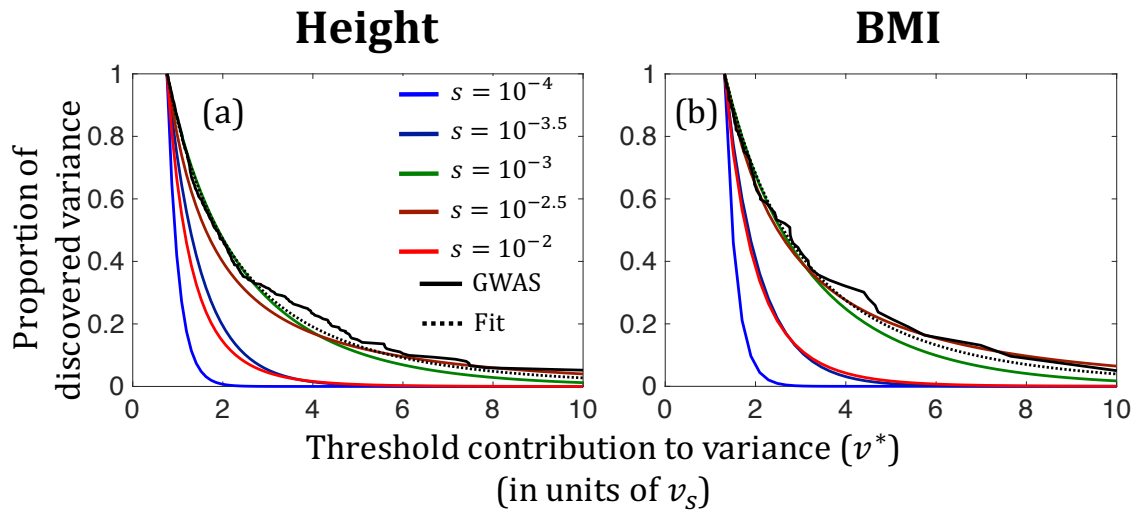

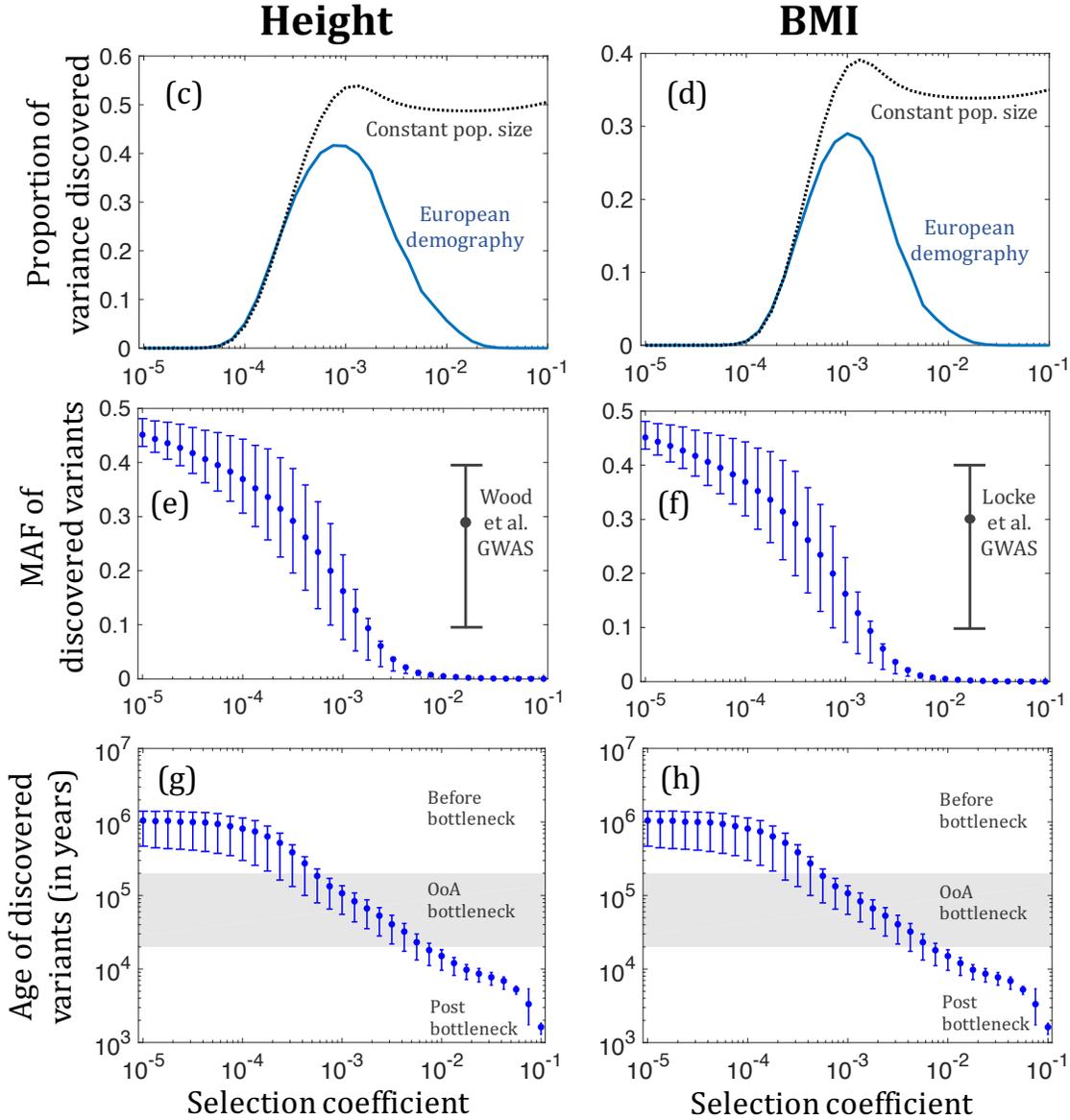

**Figure S20.** Comparison of the results of simulations with European demography with our inferences and the findings from GWAS for height and BMI (8, 10). (a & b) The distribution of variances among discovered loci. For each selection coefficient, the proportion of variance at the study threshold is set to 1. Simulation-based distributions are in color; the empirical distributions are in solid black; and our inferred fits are in dashed black. (c & d) The expected proportion of variance explained in GWAS as a function of the selection coefficient, based on simulations and on the equilibrium model with a constant population size of $N_e = 5,000$. (e & f) Comparison of MAF of discovered sites as a function of selection coefficient in simulations with the MAF observed for GWS associations in GWAS. (g & h) The age of mutations at discovered sites as a function of selection coefficient based on simulations. In (e–h), points correspond to the mean and whiskers span the 1st to 3rd quartiles of the distribution.

## 10.   The effects of genotyping

Another implication of the demographic effects that we discussed in the last section (Section S9) pertains to the reliance on genotyping rather than resequencing in GWAS. As we reviewed in Section S6, current genotyping-based GWAS typically consider only loci with MAF > 1%, for which imputation is currently quite accurate, at least in Europeans (24). Even if loci below that frequency were imputed with perfect accuracy, however, they would only be detected in a GWAS if they exceed the threshold contribution to variance for that study. Thus, loci at which the minor allele is rare would only be detected if they had very large effect sizes, which in our model implies very strong selection. For example, assuming a constant population size, if a re-sequencing study captured 25% of the heritable variance, a genotyping study with the same sample size would suffer a $\geq 50\%$ decrease in explained heritability only if $S \geq 200$. For an effective population size of $2 \cdot 10^4$ for humans (49), that implies an enormous fitness cost of $s \geq 0.5\%$ (in heterozygotes) for the minor allele. Our results for European demographic history suggest that segregating loci that are under sufficiently strong selection, and thus have sufficiently large effect sizes, to exceed current GWAS detection threshold if they had MAF just below the imputation threshold, would in fact have much lower MAF. More generally, our results suggest that there should be practically no segregating loci that fall below the current MAF imputation threshold but have sufficiently large effect sizes to exceed the variance discovery thresholds of current GWAS. This argument suggests that the common reliance on genotyping in current GWAS for quantitative traits entails practically no loss in the discovery of associations relative to resequencing.

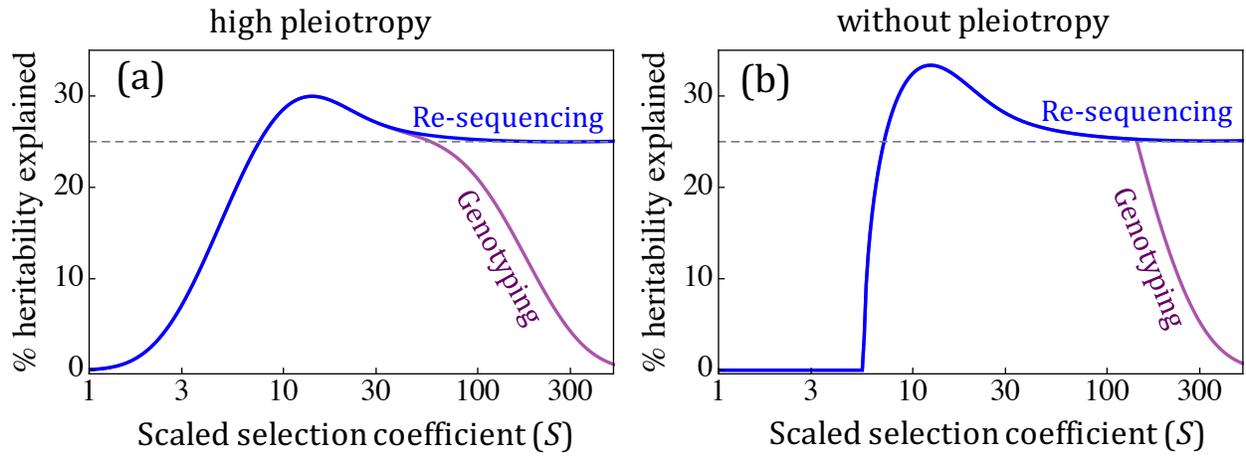

Figure S21. The heritability explained in resequencing and genotyping studies as a function of the scaled selection coefficient, assuming a constant population size, in the highly pleiotropy limit (a) and without pleiotropy (b). The study size was chosen such that a resequencing study would capture 25% of the strongly selected variance: implying a study size of $\sim 16 V_P/v_s$ in the highly pleiotropic limit (a), and a study size of $\sim 43 V_P/v_s$ without pleiotropy (b).

# 11. Glossary of notation

| Symbol | Description |
|---|---|
| $\vec{r}$ | Phenotype |
| $W(\vec{r})$ | Fitness |
| $w$ | Scale of selection |
| $n$ | Number of traits (dimensions) |
| $\vec{a}$ | $n$-dimensional effect size |
| $a_1$ | Effect size on focal trait |
| $U$ | Haploid mutation rate per generation |
| $\sigma^2$ | Phenotypic variance in a trait |
| $K$ | Number of segregating sites |
| $\varphi_n(a_1\|a)$ | Distribution of focal trait effect sizes conditional on overall effect size |
| $\eta(a_1\|S)$ | Distribution of focal trait effect sizes conditional on the scaled selection coefficient |
| $S = \dfrac{Na^2}{2w^2}$ | Scaled selection coefficient |
| $q$ | Derived allele frequency |
| $p$ | Ancestral allele frequency, $p = 1 - q$ |
| $\tau(q\|S)$ | The sojourn time for a mutation with scaled selection coefficient $S$ |
| $v$ | Contribution to variance from a site $\left(v = \tfrac{1}{2} a_1^2 q(1-q)\right)$ |
| $v_s$ | Expected contribution of a strongly selected site to variance $\left(v_s = \dfrac{2w^2}{nN}\right)$. |
| $E(V\|S)$ | Expected contribution to genetic variance from sites under selection $S$ |
| $E(K\|S)$ | Expected contribution to the number of segregating sites from sites under selection $S$ |
| $\rho(v)$ | Density of segregating sites contributing variance $v$ |

| $G(v^*)$ | Fraction of variance explained by sites contributing more than $v^*$ to the variance |
| --- | --- |
| $m$ | GWAS study size |
| H | Power to identify a locus in GWAS |

## 12.   Additional figure

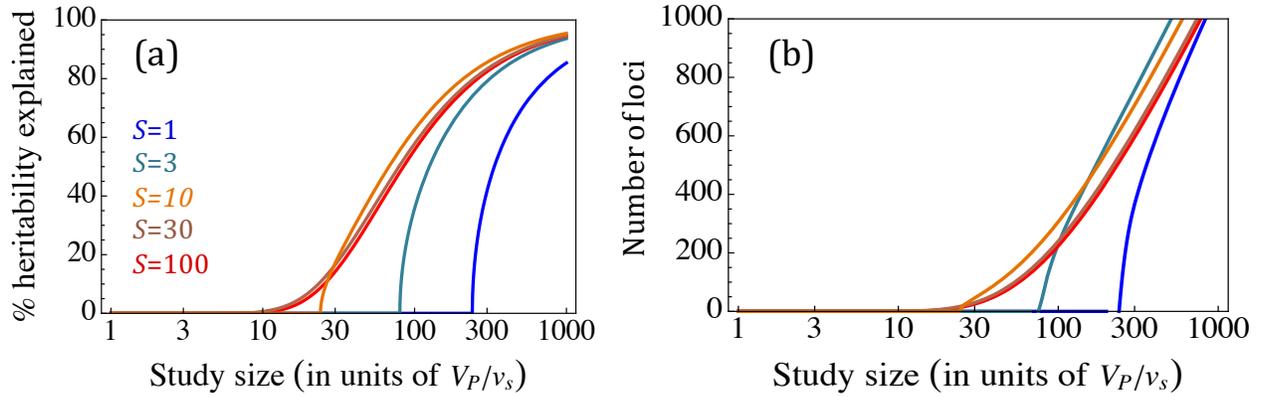

**Figure S22.** The proportion of heritability (a) and the number of variants per Mbp (b) identified in GWAS as a function of study size, in the case without pleiotropy ($n = 1$), see Section S3 for derivations. This figure is equivalent to Fig. 4 from the main text, which describes the case with pleiotropy ($n \gg 1$).